\documentclass[a4paper,fleqn]{cas-sc}
\usepackage[numbers]{natbib}
\usepackage{graphicx}
\usepackage{tikz}
\usepackage{tikz-network}
\usetikzlibrary{3d} 
\usetikzlibrary{backgrounds} 
\usetikzlibrary{shapes, arrows, positioning}
\usepackage{amsmath}
\usepackage{amssymb}
\usepackage{mathtools,xparse}
\usepackage{textcomp}
\usepackage[official]{eurosym}
\usepackage{xcolor} 
\definecolor{blue(ncs)}{RGB}{0,135,189} 
\usepackage{lipsum} 
\usepackage{siunitx}
\usepackage{algorithm}
\usepackage{algpseudocode}
\usepackage{tabularx} 
\usepackage{cleveref}
\usepackage{float}
\usepackage{subcaption} 
\usetikzlibrary{spy}
\usetikzlibrary{decorations.pathreplacing}
\usepackage{fontawesome5}

\newcommand{\xr}[1]{x_{ijbt}^{\text{r}}}
\newcommand{\xp}[1]{x_{ijbt}^{\text{p}}}
\newcommand{\xc}[1]{x_{iibt}^{\text{c}}}
\newcommand{\xvtg}[1]{x_{iibt}^{\text{V2G}}}

\DeclarePairedDelimiter{\nint}\lfloor\rceil

\DeclareMathOperator*{\minimize}{minimize}

\def\tsc#1{\csdef{#1}{\textsc{\lowercase{#1}}\xspace}}
\tsc{WGM}
\tsc{QE}
\tsc{EP}
\tsc{PMS}
\tsc{BEC}
\tsc{DE}

\begin{document}
\let\WriteBookmarks\relax
\def\floatpagepagefraction{1}
\def\textpagefraction{.001}

\shorttitle{Stochastic and Robust Optimization for EAMoD}

\shortauthors{Sten Elling Tingstad Jacobsen et~al.}

\title [mode = title]{Combined Stochastic and Robust Optimization for Electric Autonomous Mobility-on-Demand with Nested Benders Decomposition}                      

%
\author[1,2]{Sten Elling Tingstad Jacobsen}

\cormark[1]

\fnmark[1]

\ead{elling@chalmers.se}



\affiliation[1]{organization={Department of Electrical Engineering, Chalmers University of Technology},
    city={Gothenburg},
    country={Sweden}}

\author[1]{ Balázs Kulcsár}
\ead{kulcsar@chalmers.se}
\author[2]{ Anders Lindman}
\ead{anders.lindman@volvocars.com}

\affiliation[2]{organization={Volvo Cars},
    city={Gothenburg},
    country={Sweden}}

\cortext[cor1]{Corresponding author}
\cortext[cor2]{Principal corresponding author}



\begin{abstract}
The electrification and automation of mobility are reshaping how cities operate on-demand transport systems. Managing Electric Autonomous Mobility-on-Demand (EAMoD) fleets effectively requires coordinating dispatch, rebalancing, and charging decisions under multiple uncertainties, including travel demand, travel time, energy consumption, and charger availability. We address this challenge with a combined stochastic and robust model predictive control (MPC) framework. The framework integrates spatio-temporal Bayesian neural network forecasts with a multi-stage stochastic optimization model, formulated as a large-scale mixed-integer linear program. To ensure real-time applicability, we develop a tailored Nested Benders Decomposition that exploits the scenario tree structure and enables efficient parallelized solution. Stochastic optimization is employed to anticipate demand and infrastructure variability, while robust constraints on energy consumption and travel times safeguard feasibility under worst-case realizations. We evaluate the framework using high-fidelity simulations of San Francisco and Chicago. Compared with deterministic, reactive, and robust baselines, the combined stochastic and robust approach reduces median passenger waiting times by up to 36\% and 95th-percentile delays by nearly 20\%, while also lowering rebalancing distance by 27\% and electricity costs by more than 35\%. We also conduct a sensitivity analysis of battery size and vehicle efficiency, finding that energy-efficient vehicles maintain stable performance even with small batteries, whereas less efficient vehicles require larger batteries and greater infrastructure support. Our results emphasize the importance of jointly optimizing predictive control, vehicle capabilities, and infrastructure planning to enable scalable, cost-efficient EAMoD operations.
\end{abstract}


\begin{highlights}
    \item Stochastic and robust control model manages demand, energy, time, and charging uncertainties
    \item Bayesian neural networks generate multivariate travel and charging demand scenarios
    \item Scenario reduction via soft clustering preserves key patterns of future uncertainty
    \item Tailored Nested Benders decomposition efficiently solves the large-scale MILP model
    \item Simulation results show reduced costs and improved reliability over baseline methods
\end{highlights}

\begin{keywords}
Nested Benders decomposition \sep electric autonomous mobility-on-demand \sep Bayesian neural networks \sep scenario tree \sep scenario reduction \sep charging scheduling \sep connected and autonomous vehicles
\end{keywords}
\maketitle
\section{Introduction} \label{sec:intro}
As cities worldwide face rapid urban growth, environmental challenges, and evolving societal demands, transportation systems play a critical role in achieving sustainable urban development. Expanding urban populations exacerbate issues such as traffic congestion, pollution, aging infrastructure, and socioeconomic inequality, threatening the sustainability and livability of urban areas \citep{weforum, undp_urbanization}. In response, the concept of smart cities has emerged, leveraging advanced technologies such as Intelligent Transportation Systems (ITS). ITS integrate real-time data, connected infrastructure, and autonomous vehicles to enhance urban management, optimize mobility, and promote sustainable transportation solutions \citep{nature_smartcities, undp_climate, batty_smartcities}. A promising development within this domain is Electric Autonomous Mobility-on-Demand (EAMoD) systems. These systems consist of centrally coordinated fleets of electric autonomous vehicles that provide on-demand transport services \citep{booksingapore}. EAMoD systems have the potential to reduce urban emissions, improve operational efficiency, and accelerate the adoption of electric vehicles (EVs). Unlike fixed-route public transport, EAMoD offers dynamic routing and centralized coordination, enabling adaptation to real-time demand \citep{Zardini_2021}.

Optimizing such systems requires solving several interdependent problems: dispatching vehicles to satisfy spatially and temporally varying demand, rebalancing empty vehicles to anticipate future imbalances, and scheduling charging and vehicle-to-grid (V2G) operations under infrastructure and energy constraints. These decisions must be made jointly and under uncertainty, as travel demand, charger availability, travel time, and energy consumption are all stochastic and temporally correlated. Existing approaches typically address only a subset of these challenges. Many studies focus on vehicle allocation without considering the complexities introduced by electric vehicles, such as limited battery range and charging infrastructure constraints \citep{Zardini_2021}. Those that incorporate charging often treat it separately from dispatch and rebalancing, or rely on deterministic forecasts that fail to account for the full range of future uncertainty \citep{he2023data, gao2024charging}. On the scenario modeling side, constructing reduced scenario trees that faithfully represent multivariate, temporally correlated uncertainties remains computationally demanding, and most existing reduction techniques are limited to single sources of uncertainty or do not preserve the filtration structure of the stochastic process \citep{horejvsova2020evaluation}. Finally, solving the resulting multi-stage stochastic optimization problems at a scale suitable for real-time operations is itself a major challenge, as popular decomposition methods such as SDDP assume stagewise independence, an assumption violated in EAMoD applications where uncertainties exhibit strong temporal dependencies \citep{zou2019sddip}.

\begin{figure}
    \centering
    \begin{tikzpicture}[
        node distance=0.6cm and 1.5cm,
        every node/.style={align=center, font=\footnotesize},
        process/.style={rectangle, draw=black, thick, minimum width=5cm, minimum height=0.8cm, fill=white, rounded corners=10pt},
        process2/.style={rectangle, draw=black, thick, minimum width=5cm, minimum height=0.8cm, fill=white, rounded corners=10pt},
        arrow/.style={thick, ->, >=stealth},
        data/.style={rectangle, draw=black, thick, minimum width=3.5cm, minimum height=0.7cm, fill=red!20, text=black, rounded corners=10pt},
        output/.style={rectangle, draw=black, thick, minimum width=3.5cm, minimum height=0.7cm, fill=blue!20, text=black, rounded corners=10pt},
        group/.style={rectangle, draw=black, thick, rounded corners=10pt, inner sep=25pt} 
    ]

    \node (data) [data, text=black] {Historical data \\ (travel demand, charger availability, energy consumption, travel time) + current system state};

    \node (bnn_inner) [process, below=1.5cm of data,text=black] {\textbf{BNN:} Bayesian Neural Network \\ (predict demand, charging)};
    \node (treegen_inner) [process, below=0.5cm of bnn_inner,text=black] {\textbf{Scenario Tree Generation} \\ (multi-stage forecasting)};
    \node (treereduce_inner) [process, below=0.5cm of treegen_inner,text=black] {\textbf{Scenario Tree Reduction} \\ (ScenTree algorithm)};

    \node[group, fit=(bnn_inner) (treegen_inner) (treereduce_inner), color=black,below=0.5cm of data] (group1) {};

    \node at (group1.north) [anchor=north, yshift=-0.5em, font=\small\bfseries, text=black] {Prediction and Scenario Tree Details};

    \node (smpc) [process2, below=0.5cm of group1,text=black] {Scenario Model Predictive Control (SMPC)};
    \node (nestedbenders) [process2, below=0.5cm of smpc,text=black] {Nested Benders Decomposition (NBD)};

    \node (decision) [output, below=0.5cm of nestedbenders,text=black] {Fleet and charging \\ decisions};

    \draw [arrow,color=black] (data) -- (group1); 
    \draw [arrow,color=black] (bnn_inner) -- (treegen_inner);
    \draw [arrow,color=black] (treegen_inner) -- (treereduce_inner);
    \draw [arrow,color=black] (group1) -- (smpc); 
    \draw [arrow,color=black] (smpc) -- (nestedbenders);
    \draw [arrow,color=black] (nestedbenders) -- (decision);

    \end{tikzpicture}
    \caption{Overview of the scenario-based model predictive control (SMPC) framework using a Bayesian Neural Network (BNN), scenario tree generation and reduction, and Nested Benders Decomposition (NBD). Historical data is first used to train the BNN, which provides predictions of future demand and charging availability. These predictions are used to generate a multi-stage scenario tree, which is then reduced using the ScenTree algorithm to obtain a computationally tractable representation. The reduced scenario tree is used within the SMPC, which is solved using NBD to derive optimal fleet rebalancing and charging decisions.}
    \label{fig:framework}
\end{figure}
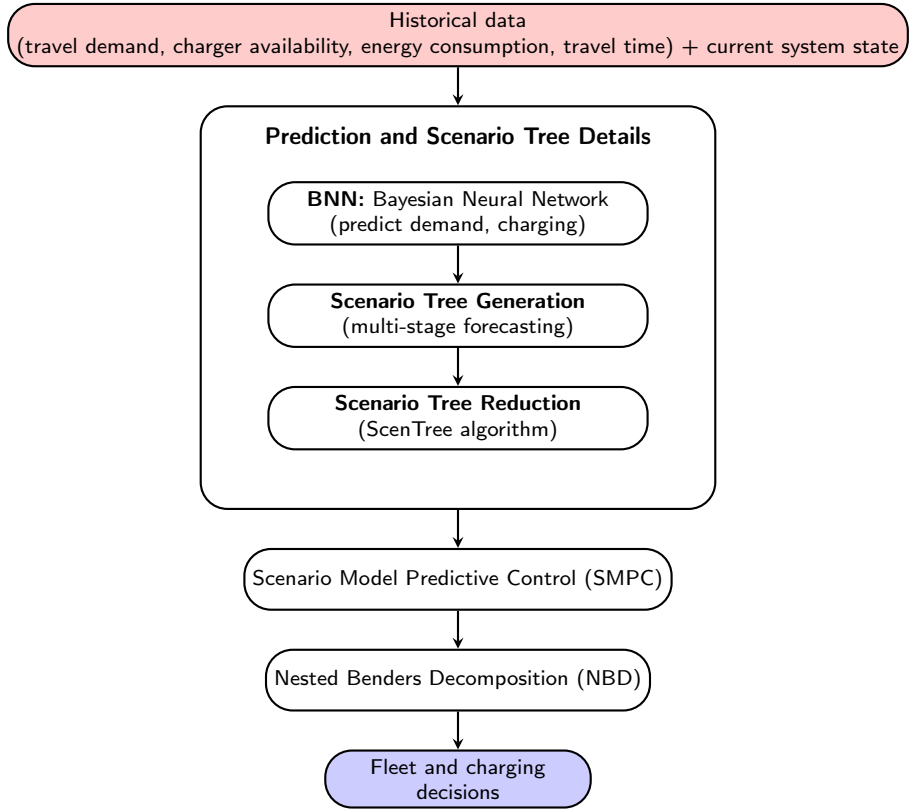

In this paper, we propose a combined stochastic and robust model predictive control (MPC) framework for EAMoD fleet optimization that addresses these challenges in an integrated manner. The framework uses Bayesian Neural Networks (BNNs) \citep{saad2024scalable} to produce calibrated probabilistic forecasts of travel demand and charger availability, capturing both spatial and temporal correlations. These forecasts are used to generate a multi-stage scenario tree, which is then reduced using a clustering-based technique that efficiently handles multiple sources of uncertainty. The resulting large-scale Mixed-Integer Linear Program (MILP) is solved using a tailored Nested Benders Decomposition (NBD) algorithm that exploits the scenario tree structure and enables parallelized computation. Robust chance constraints on energy consumption and travel time safeguard feasibility under worst-case realizations. The framework is evaluated through high-fidelity simulations in San Francisco and Chicago, using real-world taxi data, and is complemented by a comprehensive sensitivity analysis of vehicle efficiency, battery capacity, fleet size, and charging infrastructure. An overview of the framework is presented in Fig.~\ref{fig:framework}.

The main contributions of this paper are:
\begin{itemize}
   \item   A combined stochastic and robust MPC framework for EAMoD that jointly optimizes dispatch, rebalancing, charging, and V2G operations under four sources of uncertainty: travel demand, charger availability, energy consumption, and travel time.
    \item A BNN-based probabilistic forecasting pipeline integrated with a clustering-based scenario tree generation and reduction method that preserves temporal correlations across multiple uncertainty sources while maintaining computational tractability.
    \item A tailored Nested Benders Decomposition (NBD) algorithm that exploits the scenario tree structure to efficiently solve the resulting large-scale multi-stage stochastic MILP, enabling real-time applicability with parallelized subproblem solution.
    \item A comprehensive evaluation on two large-scale urban case studies (San Francisco and Chicago) demonstrating significant improvements over deterministic, reactive, and robust baselines, including up to 36\% reduction in median wait times, 20\% reduction in 95th-percentile delays, 27\% less rebalancing distance, and over 35\% savings in electricity costs.
    \item A sensitivity analysis investigating the interplay between vehicle energy efficiency, battery capacity, fleet size, and charger availability, providing practical insights for fleet design and infrastructure planning in EAMoD systems.
\end{itemize}
Importantly, the proposed framework differs from existing multi-stage 
approaches in two key aspects. First, the uncertainty process is 
temporally dependent and multivariate, violating the stagewise 
independence assumption required by SDDP-type methods. Second, the 
integration of robust chance constraints within a multi-stage SMPC 
architecture enables feasibility guarantees while retaining 
anticipatory decision-making. These features necessitate a tailored 
Nested Benders Decomposition capable of handling large scenario trees 
with temporal coupling.

The remainder of this work is organized as follows. Section~\ref{sec:relwork} reviews the related literature on EAMoD optimization, scenario tree methods, and decomposition algorithms. Section~\ref{section:model} presents the system model and the complete multi-stage robust and stochastic MPC formulation. Section~\ref{section:scentree} describes the BNN-based scenario generation and clustering-based reduction procedure. Section~\ref{sec:NB} presents the NBD algorithm. Section~\ref{sec:sim} reports simulation results for San Francisco and Chicago, including a sensitivity analysis, and concludes with a discussion of findings and future research directions.
\newpage
\section{Related Work}\label{sec:relwork}

\subsection{AMoD demand modeling and uncertainty quantification}

Optimizing EAMoD systems introduces several challenges, particularly in addressing demand-supply imbalances resulting from spatial and temporal variations in travel demand.  Numerous studies have proposed various vehicle allocation control methods to mitigate these imbalances. For instance, \citep{zhang2016model} introduced a Model Predictive Control (MPC) approach that demonstrated promising results; however, its scalability was limited due to the need to model the entire road network. To enhance scalability, many studies have adopted discretization strategies, dividing the city into larger areas, often referred to as stations. An example of this approach is presented in \citep{iglesias2019bcmp}, where a queue-theoretical model was employed. An alternative and effective model-free approach is the "+1 method," which directs vehicles to the most recent service request location, as demonstrated by \citep{ruch2020the+}. For MPC methods, incorporating uncertainties into the control algorithm has proven important. A stochastic MPC framework combined with a Long Short-Term Memory (LSTM) neural network for travel demand prediction, leveraging Sample Average Approximation (SAA) to model uncertainty, was proposed by \citep{tsao2018stochastic}. Similarly, \citep{jacobsen2023predictive} introduced a chance-constrained approach using Gaussian Process Regression (GPR) for demand prediction, while \citep{miao2021data} proposed a data-driven distributionally robust optimization method. Despite these advancements, many existing strategies often overlook the complexities associated with electric vehicles (EVs), including limited battery capacities and charging requirements, as highlighted in the review by \citep{Zardini_2021}.

To address this gap, several studies have explored charge scheduling strategies in EAMoD. Charging strategies such as plug-in versus battery swapping have been investigated to optimize fleet availability and operational efficiency in EAMoD systems \citep{gao2024charging}. The placement of charging stations using stochastic programming and considering different uncertainties was studied in \citep{faridimehr2018stochastic}. The charging dynamics based on data from an EV taxi fleet were examined in \citep{lei2022understanding}. Fleet sizing and charging station placement considering endogenous congestion were analyzed in \citep{yang2023fleet}. While these studies address important aspects of EAMoD fleets, the challenging task of simultaneously considering charge scheduling and vehicle allocation is not explored. Some of these aspects are considered in \citep{he2023data}, where a distributionally robust approach was proposed to account for uncertainties in demand and supply. An ARIMA model was used for travel demand prediction, and the uncertainty set was constructed using bootstrapping. Different reinforcement learning algorithms have also been proposed. A queue-theoretical deep reinforcement learning method incorporating routing, battery charging, and dynamic pricing was introduced by \citep{turan2020dynamic}. Various reinforcement learning strategies are compared in the survey by \citep{qin2022reinforcement}. However, a comprehensive approach that simultaneously addresses both charging scheduling and vehicle allocation while accounting for uncertainties in travel demand, charging availability, energy consumption and travel time remains unexplored.

\subsection{Scenario generation, reduction, and quality metrics}

A common approach to handling uncertainties in control and optimization is to construct scenario trees. A scenario tree provides a structured representation of the evolution of uncertainty over time in a multi-stage decision process. Each node in the tree corresponds to a possible future state, while the branches represent different ways in which uncertainty can unfold, along with their associated probabilities. It has been demonstrated that increasing the number of scenarios in the tree improves the accuracy of the uncertainty representation. However, it should be noted that an increase in the number of scenarios also results in an increase in the computational complexity of the optimization problem \citep{beltran2021two}. In order to address this issue, a range of scenario reduction techniques have been developed with the objective of constructing a reduced scenario tree that closely approximates the original one. These methods typically rely on a metric that quantifies the difference between the original and reduced trees, which can be used both to guide the reduction procedure and to evaluate its quality. A prevalent methodology in this field is moment matching, whereby the reduced scenario tree is constructed in order to preserve the first few moments (e.g., mean, variance, skewness, and kurtosis) of the original stochastic process \citep{hoyland2001generating}. This technique has been employed in several studies to generate scenario trees by matching the first four moments  \citep{ansaripoor2016recursive,bao2023learning}. However, it has been shown that two probability distributions with matching moments can still be significantly different \citep{hochreiter2007financial}. An alternative metric is the Wasserstein distance, which quantifies the similarity between two probability distributions and has been used for scenario reduction \citep{dupavcova2003scenario}. However, the Wasserstein distance does not consider the filtration structure of the scenario tree, meaning it does not account for the temporal correlations between different stages. To address this, the Nested distance, also referred to as process distance, was introduced in \citep{pflug2010version}. The Nested distance explicitly accounts for the tree structure, preserving the temporal dependencies of the stochastic process. Although computing the Nested distance between two scenario trees is relatively straightforward, finding an optimally reduced scenario tree that minimizes this distance is computationally challenging. One approach to scenario reduction based on the Nested distance involves multistage k-means clustering, as proposed in \citep{kovacevic2015tree}. However, this method still requires solving multiple large-scale linear programs (LPs), making it computationally demanding. A comprehensive evaluation of different scenario reduction techniques using the Nested distance is provided in \citep{horejvsova2020evaluation}, which highlights that methods minimizing the Nested distance tend to be computationally expensive. Another challenge is reducing scenario trees that incorporate multiple sources of uncertainty.

\subsection{Multistage stochastic optimization and decomposition methods}

Despite advances in scenario reduction, solving multi-stage stochastic programs (SPs) remains computationally challenging, particularly as the number of scenarios and the planning horizon increase. To address this, a wide range of decomposition techniques have been developed. The classical Benders decomposition method~\citep{benders1962partitioning} laid the foundation for solving mixed integer linear programs (MILPs) with complicating variables, and was extended into the L-shaped method ~\citep{vanslyke1969} for two-stage SPs. Subsequent work introduced regularized variants~\citep{ruszczynski1986regularized} and stochastic decomposition methods~\citep{higle1991stochastic,higle1994framework} inspired by stochastic quasi-gradient approaches~\citep{ermol1983stochastic} and stochastic approximation~\citep{robbins1951stochastic}. For multistage problems, nested Benders decomposition (NBD)~\citep{birge1985decomposition} and multistage stochastic decomposition~\citep{sen2014multistage} were developed to extend these ideas. A major advancement came with sampling-based methods such as stochastic dual dynamic programming (SDDP)~\citep{pereira1991multi}, and its subsequent enhancements~\citep{infanger1996multi,shapiro2011analysis}, as well as the stochastic dual dynamic integer programming (SDDiP) approach~\citep{zou2019sddip}. Alternative strategies include abridged nested Benders~\citep{donohue2006abridged} and cutting-plane or partial sampling methods~\citep{chen1999parallel}. SDDP and SDDiP are especially effective for convex or mixed-integer  multistage problems but typically rely on the assumption of stagewise  independence of the underlying uncertainty process, which facilitates  cut sharing across nodes at the same stage and enables tractable  backward recursion~\citep{shapiro2011analysis,vanackooij2017conditional}.  While extensions exist that embed dependence via additional state  variables, the stagewise independence assumption remains central to  the classical framework. This assumption can be restrictive in  real-world applications where uncertainties exhibit strong temporal  correlations. In EAMoD systems, travel demand, energy consumption,  and charger availability evolve with significant temporal dependence,  violating the structural assumptions underlying standard SDDP-type  methods and motivating alternative decomposition strategies that  explicitly operate on temporally dependent scenario trees.

In transportation research, variants of Benders decomposition have been successfully applied to a variety of different problems. A stabilized Benders algorithm has been used to solve a two-stage stochastic program for locating recharging stations in one-way electric car-sharing systems under demand uncertainty, demonstrating scalability to hundreds of large-scale demand scenarios \citep{CALIK2019121}. An exact Benders decomposition embedded in a branch-and-cut framework has been developed to strategically locate mini-depots for last-mile crowd-shipping, achieving significant computational speedups and cost savings \citep{NIETOISAZA202262}. A hybrid approach combining Benders decomposition with NSGA-II has been proposed for the joint optimization of EV charging station placement and pricing, enabling efficient infrastructure deployment while balancing conflicting objectives such as consumer pricing and operator profitability \citep{AMEER2025126385}.Despite these advances, existing decomposition methods either assume 
stagewise independence (e.g., SDDP), rely on convexity, or are not 
designed for real-time MPC implementations with integer variables and 
temporally correlated uncertainties. The EAMoD setting considered here 
requires solving large-scale multi-stage MILPs with coupled 
uncertainties and integrality constraints, motivating the development 
of a tailored Nested Benders approach.

The fleet composition, including the fleet size and individual vehicle attributes such as battery size and energy efficiency, affects how the fleet is optimized, the service quality provided, and the operational cost. For example, a less energy-efficient vehicle would need to charge more often compared to an efficient vehicle. Alternatively, the less energy-efficient vehicle could have a larger battery to compensate for its inefficiency, but this would increase costs and have a higher environmental impact. Few studies have incorporated fleet composition into the evaluation of EAMoD fleet performance. The effects of battery size and fleet size have been explored in \citep{paparella2023electric}. However, fast charging and varying energy efficiency were not considered.

\section{EAMoD vehicle movement modeling} \label{section:model}

The model presented in this section is designed for centralized control of electric mobility-on-demand systems, focusing on optimizing vehicle movement and bi-directional charging scheduling to minimize costs while maximizing service completion levels. These high-level decisions serve as references for low-level controllers, which make real-time adjustments. Central to our model is its predictive nature across time, space, and energy dimensions. By exploiting its predictive capabilities, we aim to make optimal decisions in the current time step based on expected future states evolving in time and space. The model is tailored for receding horizon optimization, ensuring adaptability to changing conditions. 

Our model is comprised of two integral components. First, we build upon the network model previously established in \citep{jacobsen2023predictive} by incorporating electric vehicles (EVs) into the framework. This enhancement requires adding several key features specific to EVs. First, charging stations to accommodate the recharging of vehicles, second the limited range of EVs necessitating strategic planning around vehicle deployment, and third vehicle-to-grid (V2G) capabilities, which allow for the interaction between the vehicle's battery and the power grid for energy exchange. Secondly, we introduce uncertainties into the model, encompassing aspects such as demand variability, charging station availability, travel time, and energy consumption rates.

\subsection{Model Notation}
 The Electric Autonomous Mobility on Demand (EAMoD) framework represents the city as comprising $n$ distinct stations or areas, enumerated as $\mathcal{N} = [1, \ldots, n]$. Stations are identified by applying a K-means clustering algorithm to travel data. The chargers, positioned at each station's centroid, are publicly accessible and the charger occupancy is assumed to follow a non-stationary Poisson process. This simplification avoids the complexities of optimizing charger locations, for which existing literature provides insights \citep{kchaou2021charging,yang2017data}. Travel is characterized by origin-destination pairs, with the origin denoted by index $i \in \mathcal{N}$ and the destination by index $j \in \mathcal{N}$. The state of charge (SoC) of vehicles is categorized into a set $\mathcal{B} = [0, 0.01, \ldots, 0.99, 1]$, defining each SoC level by the index $b$. Time is discretized into $\mathcal{T} = [1, 2, \ldots, T]$, where $t$ is the time index and $T$ the prediction horizon. A summary of different variables used in the model is provided in Table \ref{tab:var_combined}.
 
\begin{table}
\caption{Summary of all indices, state variables, decision variables, slack variables, and parameters used in the model.}
\label{tab:var_combined}
\begin{tabular}{lll}
\hline
\textbf{Category} & \textbf{Symbol} & \textbf{Description} \\ \hline
\multicolumn{3}{l}{\textbf{Indices}} \\ 
 & $i \in \mathcal{N}$ & Origin of travel  \\
 & $j\in \mathcal{N}$ & Destination of travel  \\
 & $b\in \mathcal{B}$ & SoC  \\
 & $t \in \mathcal{T}$ & Time  \\
 & $s \in \mathcal{S}$ & Scenario sample  \\ \hline

\multicolumn{3}{l}{\textbf{State Variables}} \\ 
 & $\lambda_{ij} \in \mathbb{N}$ & Number of passengers traveling from $i$ to $j$ \\
 & $\phi_{ibt}\in \mathbb{N}$ & Number of vehicles becoming available at station $i$ with SoC $b$ at time $t$ (exogenous) \\
 & $\text{k}^{\text{c}}_{it}\in \mathbb{N}$ & Number of available chargers at station $i$ at time $t$  \\ \hline

\multicolumn{3}{l}{\textbf{Decision Variables}} \\ 
 & $x_{ibt}^{\text{c}}\in \mathbb{N}$ & Vehicles charged at station $i$ with SoC $b$  \\
 & $x_{ibt}^{\text{d}}\in \mathbb{N}$ & Vehicles used for V2G at station $i$ with SoC $b$  \\
 & $x_{ijbt}^{\text{p}}\in \mathbb{N}$ & Vehicles transporting passengers from $i$ to $j$ with SoC $b$ at time $t$  \\
 & $x_{ijbt}^{\text{r}}\in \mathbb{N}$ & Vehicles rebalancing from $i$ to $j$ with SoC $b$ at time $t$  \\ \hline

\multicolumn{3}{l}{\textbf{Slack Variables}} \\ 
 & $s_{ijt} \in \mathbb{N}_{\geq 0}$ & Customers not served at station $i$ wishing to travel to $j$ at time $t$ (non-negative)  \\ \hline

\multicolumn{3}{l}{\textbf{Parameters}} \\ 
 & $\Delta b^{\text{c}}$ & SoC charge in one time interval \\
 & $\Delta \text{d}$ & SoC discharge to the grid in one time interval \\
 & $\text{e}^\text{b}$ & Battery capacity [kWh] \\
 & $e_{ijt}$ & Energy consumption for travel from $i$ to $j$ at time $t$ \\
 & $P^{\text{c}}_b$ & Charging power in kWh for different SoC levels $b$ \\
 & $P^{\text{d}}$ & V2G transfer power in kWh \\
 & $\Delta t_{ijt}$ & Travel time between station $i$ and $j$ at time $t$ \\
 & $\Delta b_{ijt}$ & SoC consumption between station $i$ and $j$ at time $t$ \\
 & $b_{ij}^{\text{min}}$ & Minimum required SoC to drive a customer from station $i$ and $j$ at time $t$ \\
 & $s$ & Index for different scenarios \\
 & $S_t$ & Set of scenarios at stage $t$ \\
 & $b_{\text{max}}$ & Max SoC threshold that is penalized in the terminal cost  \\
 & $q_{\text{max}}$ & Maximum penalty in terminal cost \\ 
 & $e_{\text{safety}}$ & Safety buffer for energy consumption [kWh] \\
 & $\hat{\tau}_{ijt}$ & Conservative travel time estimate for route $(i,j)$ at time $t$ \\ \hline
\end{tabular}
\end{table}

\subsection{Energy, Space and Time Network Model}
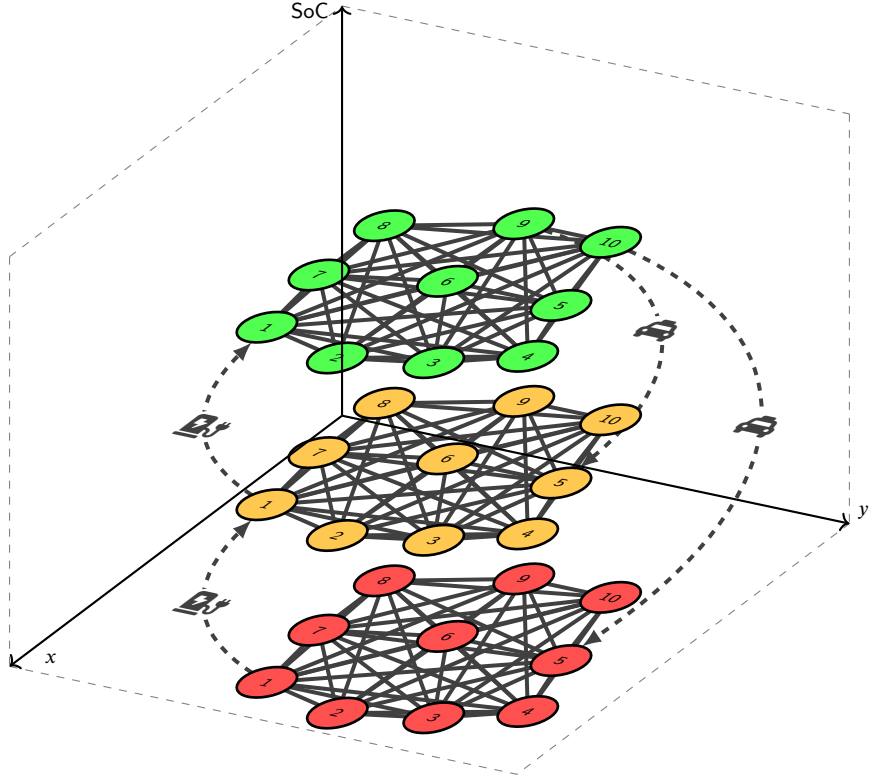
\begin{figure}
\centering
\begin{tikzpicture}[multilayer=3d,scale=0.9]
\SetVertexStyle[MinSize=.7cm]
\SetDefaultUnit{cm}

\SetEdgeStyle{color=black, thick, opacity=0.5}
\SetLayerDistance{-2.6}
\definecolor{soc1}{RGB}{255,80,80}     
\definecolor{soc2}{RGB}{255,200,80}    
\definecolor{soc3}{RGB}{80,255,80}     

\def\xcoords{{0.8,2.1,3.3,4.3,3.9,2.2,0.6,0.6,2.2,3.5}}
\def\ycoords{{0.4,0.1,0.4,0.9,2.0,2.0,1.6,2.8,3.4,3.4}}

\foreach \i in {1,...,10} {
    \pgfmathparse{\xcoords[\i-1]} \let\x\pgfmathresult 
    \pgfmathparse{\ycoords[\i-1]} \let\y\pgfmathresult 
  \Vertex[x=\x, y=\y, label=\i, layer=1,color=soc3]{v\i}
}
\foreach \i in {1,...,10} {
  \foreach \j in {1,...,10} {
    \ifnum\j>\i
      \Edge(v\i)(v\j) 
    \fi
  }
}

\foreach \i in {1,...,10} {
  \pgfmathparse{\xcoords[\i-1]} \let\x\pgfmathresult
  \pgfmathparse{\ycoords[\i-1]} \let\y\pgfmathresult
  \Vertex[x=\x, y=\y, label=\i, layer=2,color=soc2]{w\i}
}
\foreach \i in {1,...,10} {
  \foreach \j in {1,...,10} {
    \ifnum\j>\i
      \Edge(w\i)(w\j) 
    \fi
  }
}

\foreach \i in {1,...,10} {
  \pgfmathparse{\xcoords[\i-1]} \let\x\pgfmathresult
  \pgfmathparse{\ycoords[\i-1]} \let\y\pgfmathresult
  \Vertex[x=\x, y=\y, label=\i, layer=3,color=soc1]{z\i}
}
\foreach \i in {1,...,10} {
  \foreach \j in {1,...,10} {
    \ifnum\j>\i
      \Edge(z\i)(z\j) 
    \fi
  }
}

\Edge[Direct,bend=30,style={dashed},label={\Large \faIcon[solid]{charging-station}}](w1)(v1)
\Edge[Direct,bend=30,style={dashed},label={\Large \faIcon[solid]{charging-station}}](z1)(w1)
\Edge[Direct,bend=40,style={dashed},label={\Large \faIcon[solid]{taxi}}](v10)(z5)
\Edge[Direct,bend=40,style={dashed},label={\Large \faIcon[solid]{taxi}}](v9)(w5)
\SetEdgeStyle{color=black, thick, opacity=0.5}
\def\xmin{0.7}
\def\xmax{6.8}
\def\ymin{-3.3}
\def\ymax{4.3}
\def\zmin{0}         
\def\zmax{-6.0}      

\begin{scope}[canvas is xy plane at z=\zmax]
  \draw[dashed] (\xmin,\ymin) rectangle (\xmax,\ymax);
\end{scope}

\begin{scope}[canvas is xy plane at z=\zmin]
  \draw[dashed] (\xmax,\ymin) -- (\xmax,\ymax);     
  \draw[dashed] (\xmin,\ymin) -- (\xmax,\ymin);   
\end{scope}

\foreach \x/\y in {\xmax/\ymin, \xmax/\ymax, \xmin/\ymin} {
  \draw[dashed] (\x,\y,\zmin) -- (\x,\y,\zmax);
}
\node[text=black] at (\xmin+0.5,\ymin,\zmax) [below right] {$x$};    
\node[text=black] at (\xmax,\ymax+0.2,\zmax) [above] {$y$};          
\node[text=black] at (\xmax-0.1,\ymin,\zmin) [left] {SoC};           

\draw[->, thick, black] (\xmax,\ymin,\zmax) -- (\xmin,\ymin,\zmax) ; 
\draw[->, thick, black] (\xmax,\ymin,\zmax) -- (\xmax,\ymax,\zmax); 
\draw[->, thick, black] (\xmax,\ymin,\zmax) -- (\xmax,\ymin,\zmin); 
\end{tikzpicture}
\caption{This diagram illustrates a multi-layer network representing an Electric Autonomous Mobility-on-Demand (EAMoD) system. Each layer corresponds to a different battery state of charge (SoC). When a vehicle travels in the network (illustrated by a dashed arrow with a vehicle icon), it consumes energy, leading to a decrease in its state of charge (SoC). In contrast, a dashed arrow with a charging station represents a charging trip, during which the vehicle remains at the station and its SoC increases.}
\label{fig:roadnetwork}
\end{figure}

To accurately represent the dynamic environment in which vehicles operate, we introduce an Energy, Space, and Time Network Model. This model forecasts the number of vehicles at the different stations in different time intervals, and their SoC levels, providing a comprehensive understanding of fleet operations (see Fig. \ref{fig:roadnetwork}). Rather than tracking individual vehicle trajectories, the model aggregates vehicles within each station based on identical states. This aggregation enhances computational tractability, enabling efficient analysis of fleet distribution and status across the network while maintaining manageable complexity. The stations are assumed to be fully interconnected, allowing unrestricted travel between any two locations. Consequently, the road network is modeled as a complete graph, ensuring total connectivity among all stations. In this framework, the travel time and energy consumption between any two stations $i$ and $j$ at time $t$ are denoted as $\Delta t_{ijt}$ and $\Delta b_{ijt}$, respectively. These values are time-dependent, as they may vary throughout the day.

Uncertainty at stage $t$ is represented by the random vector $\xi_t$, with history 
$\xi_{[t]} = (\xi_1,\dots,\xi_t)$. In the scenario tree approximation, each scenario 
is indexed by $s \in \mathcal{S}$, and the corresponding realization of uncertainty 
at stage $t$ is denoted $\xi_t^s$. Decision variables are therefore indexed by $s$, 
while $\xi_t$ is reserved exclusively for the underlying random variable. 
This distinction ensures a clear separation between the stochastic process and its 
sampled realizations in the Sample Average Approximation (SAA) formulation.

\subsubsection{Flow dynamics}\label{sec:flowdyn}
There are three types of vehicle movements in the network model. First, passenger transport is represented by \( x^{\text{p}}_{ijbt} \), denoting vehicles traveling from origin \( i \) to destination \( j \) with SoC \( b \) at time \( t \). Second, vehicle relocation is modeled using \( x^{\text{r}}_{ijbt} \), which represents vehicles moving between stations without passengers, a crucial mechanism for balancing vehicle distribution across the network. Third, vehicles engaged in charging or V2G operations are represented by \( x^{\text{c}}_{iibt} \) and \( x^{\text{d}}_{iibt} \), respectively. The index \( ii \) is used for these operations because charging and V2G are modeled as trips within the same station. 

The flow conservation constraint tracks vehicle movements across the network over time. At each time step, vehicles may become available at a station from two sources: (i)~vehicles that were stationary at the start of the optimization horizon ($t = 0$), representing the initial fleet distribution, and (ii)~vehicles that were dispatched before the current horizon began and arrive during the planning period ($t > 0$). We denote by $\phi_{ibt}$ the number of vehicles becoming available at station~$i$ with SoC~$b$ at time~$t$. This is an exogenous parameter determined by the current system state: at $t = 0$, it reflects the initial fleet distribution; at $t > 0$, it captures predetermined arrivals from trips initiated in previous receding horizon iterations. These arrivals cannot be altered by the current optimization.

To ensure conservation of vehicles in the system, we enforce a flow conservation constraint at each station:

\begin{equation}
    \begin{aligned}[t]\label{eq:flowcons} 
            &x_{iibt}^{\text{c}} + x_{iibt}^{\text{d}}  
             + \sum_{j=1}^\mathcal{N} \left [ x_{ijbt}^{\text{p}} + x_{ijbt}^{\text{r}} \right ]  = \phi_{ibt} + x_{ii(b-\Delta b^{\text{c}})(t-1)}^{\text{c}} + x_{ii(b+\Delta d)(t-1)}^{\text{d}} + \sum_{j=1}^\mathcal{N} \left [ 
            x_{ji(b+\Delta b_{jit})(t-\Delta t_{jit})}^{\text{p}} + x_{ji(b+\Delta b_{jit})(t-\Delta t_{jit})}^{\text{r}} \right ] \\
            & \qquad \forall i \in \mathcal{N},\ t \in \mathcal{T},\ \forall b \in \mathcal{B}.
        \end{aligned}  \\
\end{equation}

where \( \Delta b_{jit} \) denotes the expected SoC reduction during a trip from station \( j \) to station \( i \), and \( \Delta b^{\text{c}} \) and \( \Delta d \) represent energy gained or lost through charging and V2G operations, respectively. The left-hand side represents the outflow of vehicles from station i and the right-hand side the inflow of vehicles to station i. $\phi_{ibt}$ is the initial condition of the inflow from $t<0$. At the start of each receding horizon iteration, $\phi_{ibt}$ is computed from the known positions and scheduled arrival times of all vehicles in the fleet.

The first four terms in Eq. \eqref{eq:flowcons} represent vehicles entering or exiting charging or V2G services. The summation terms account for vehicles departing (first two terms) and arriving (last two terms) at each station. This equation ensures system-wide vehicle conservation while incorporating temporal and energy constraints. Specifically, each vehicle movement is subject to a travel time delay \( \Delta t_{jit} \) and an associated SoC reduction \( \Delta b_{ijt} \), given by:
\begin{equation*}
\Delta b_{ijt} = \nint{\frac{e_{ijt} }{e^{\text{b}}}},
\end{equation*}
where \( e^{\text{b}} \) is the battery capacity and \( e_{ijt} \) is the energy consumption for travel between stations \( i \) and \( j \) at time \( t \).
Since energy consumption and travel time are subject to uncertainty (e.g., due to traffic conditions, auxiliary loads, and route variability), the parameters $\Delta b_{ijt}$ and $\Delta t_{jit}$ used in the flow conservation constraint~\eqref{eq:flowcons} must be chosen conservatively to ensure operational feasibility. We determine these values using chance constraints on the underlying random variables. Specifically, the travel time parameter is set to a conservative estimate $\hat{\tau}_{ijt}$ satisfying
\begin{equation}\label{eq:cctime_model}
\Delta t_{jit} = \hat{\tau}_{jit}, \quad \text{where } \mathbb{P}(\tau_{jit} \leq \hat{\tau}_{jit}) \geq 1 - \epsilon^{\text{t}},
\end{equation}
ensuring that the actual travel time does not exceed the value used in the model with probability at least $1 - \epsilon^{\text{t}}$. Similarly, the energy consumption parameter is set to
\begin{equation}\label{eq:ccenergy_model}
\Delta b_{ijt} = \hat{b}_{ijt}, \quad \text{where } \mathbb{P}\!\left(e_{ijt} \leq \hat{b}_{ijt} \cdot e^{\text{b}} + e_{\text{safety}}\right) \geq 1 - \epsilon^{\text{b}},
\end{equation}
where $e_{\text{safety}}$ is a safety buffer. These conservative estimates are computed offline at the start of each receding horizon iteration using the empirical distributions of travel time and energy consumption, and are then treated as fixed parameters in the optimization. Smaller values of $\epsilon^{\text{b}}$ and $\epsilon^{\text{t}}$ correspond to more conservative (robust) policies.

\subsubsection{Imbalance dynamics}
EAMoD systems are prone to imbalances caused by spatial and temporal variations in travel demand. To mitigate these imbalances through effective vehicle allocation, it is essential to model the imbalance at each station. We define the imbalance at a station as the difference between the number of passengers awaiting service and the number of available vehicles. Ideally, system operations should maintain this imbalance at zero. However, due to demand fluctuations and operational constraints, achieving a perfect balance is often infeasible. Therefore, we introduce the following constraint, which allows for a positive imbalance:
\begin{equation}\label{eq:im} \lambda_{ijt} - \sum_{b=b_{ij}^{\min}}^{1} x^p_{ijbt} + s_{ij(t-1)} = s_{ijt}, \quad \forall i, j \in \mathcal{N}, \forall t \in \mathcal{T}, \end{equation}
where $\lambda_{ijt}$ denotes the travel demand from station $i$ to station $j$ at time $t$, and $x^p_{ijbt}$ represents the number of vehicles with a SoC between $b_{ij}^{\min}$ and 1. The threshold $b_{ij}^{\min}$ is the minimum SoC required for a vehicle to complete a trip from $i$ to $j$ with high probability, derived from the energy consumption chance constraint~\eqref{eq:ccenergy_model}:
\begin{equation}\label{eq:bmin_def}
b_{ij}^{\min} = \min\!\left\{b \in \mathcal{B} : \mathbb{P}\!\left(b \cdot e^{\text{b}} - e_{ijt} \geq e_{\text{safety}}\right) \geq 1 - \epsilon^{\text{b}}\right\}.
\end{equation}
This ensures that only vehicles with sufficient charge to complete the trip (including a safety buffer $e_{\text{safety}}$) with probability at least $1 - \epsilon^{\text{b}}$ are eligible for dispatch. By restricting the summation in~\eqref{eq:im} to SoC levels $b \geq b_{ij}^{\min}$, the constraint directly links vehicle dispatch decisions to the robust energy requirements. The variable $s_{ijt}$ is a slack variable that captures unmet demand, thereby allowing some flexibility in maintaining balance.   


\subsubsection{Charging Dynamics}

Charging dynamics within the network describe how the SoC of vehicles at each charging node evolves. The increase in SoC, denoted as $\Delta b$, during each time interval is calculated by the equation:
\begin{equation*}
\Delta b^{\text{c}} = \left\lceil{\frac{P^{\text{c}}(b) \Delta t}{e^b}}\right\rceil.
\end{equation*}
In this equation, $P^{\text{c}}(b)$ is the SoC-dependent charging power and $\Delta t$ is the discrete time-step length. The operator $\left\lceil \cdot \right\rceil$ rounds $\Delta b^{\text{c}}$ up to the nearest integer, ensuring discrete SoC bins. Charging power typically exhibits non-linear behavior with SoC rates slow as the battery approaches full charge. We adopt the charging-power curve $P^{\text{c}}(b)$ from \citep{WASSILIADIS}, which remains approximately constant up to an SoC of $45\%$ and then decreases linearly.

Within each discrete time interval, vehicles at a charging station may begin a charging session, continue a session from a previous interval, or complete their charging. We assume a one-interval, $\Delta t$, charging duration per decision step. This simplification fits with the receding horizon control strategy used in the model, where decisions are regularly updated. By limiting charging sessions to a single interval, the model remains effective and responsive to changes in charging demand and infrastructure availability, without the need to manage the complexities of prolonged charging durations.

The energy depletion of the battery when the vehicle is V2G is given by:
\begin{equation*}
\Delta d = \left\lceil{\frac{P^{\text{d}} \Delta t}{e^b}}\right\rceil.
\end{equation*}
where $P^{\text{d}}$ is the power delivered to the grid, which is considered to be constant. Additionally, the availability of chargers at each station is limited, which necessitates a constraint on the total number of vehicles that can charge or await charging at any given time:

\begin{equation}\label{eq:charger}
\sum_b x_{ibt}^{\text{c}} + x_{ibt}^{\text{d}} \leq \text{k}^{\text{c}}_{it}, \quad \forall i \in \mathcal{N}, \forall t \in \mathcal{T},
\end{equation}
where $\text{k}^{\text{c}}_{it}$ represents the total number of charging stations available at each site. This constraint ensures that the model accurately reflects the actual limitations of the charging infrastructure, preventing over-allocation of charging resources.

\subsection{Objective Function}

We formulate an economic objective function in which each term corresponds 
to a specific monetary cost or revenue. The goal is to minimize the expected 
total cost under uncertainty in travel demand, charging-station occupancy, 
energy consumption, and travel time. The true, unknown, and time-varying 
probability distributions of these random variables are denoted by
\(
\mathcal{D}_{\lambda}^*,\ 
\mathcal{D}_{k}^*,\ 
\mathcal{D}_{b}^*,\ \text{and }\ 
\mathcal{D}_{\tau}^*,
\)
respectively. The uncertainties in energy consumption $\mathcal{D}_{b}^*$ 
and travel time $\mathcal{D}_{\tau}^*$ are handled via chance constraints. 
The distributions $\mathcal{D}_{\lambda}^*$ and $\mathcal{D}_{k}^*$ enter 
directly into the expected cost. At each stage $t \in \{1,\ldots,T\}$, the exogenous uncertainty is 
described by the joint random vector
\begin{equation}
    \xi_t \triangleq \bigl(\{\lambda_{ijt}\}_{i,j=1}^{N},\; 
    \{k_{it}^{c}\}_{i=1}^{N}\bigr),
\end{equation}
where $\lambda_{ijt} \geq 0$ is the travel demand from zone $i$ to zone 
$j$ at stage $t$, and $k_{it}^{c} \geq 0$ is the charging-station 
occupancy at zone $i$ at stage $t$. We denote the history of observations 
up to stage $t$ by $\xi_{[t]} \triangleq (\xi_1,\ldots,\xi_t)$. The 
joint conditional distribution of $\xi_t$ given the history $\xi_{[t-1]}$ 
is denoted $\mathcal{D}_t(\cdot \mid \xi_{[t-1]})$.

The multi-stage SMPC objective is:
\begin{align}
\label{eq:smpcobjective}
J &= \underbrace{\sum_{i=1}^{N}\sum_{j=1}^{N}\sum_{b=0}^{1} 
      Q\!\left(x_{ijb0}^{\mathrm{r}},\, x_{ijb0}^{\mathrm{p}},\,
               x_{iib0}^{\mathrm{c}},\, x_{iib0}^{\mathrm{d}}\right)
    }_{\text{stage } t=0}
 + \mathbb{E}_{\xi_1 \sim \mathcal{D}_1}\Bigg[
    \sum_{i,j,b} Q_1
    + \mathbb{E}_{\xi_2 \sim \mathcal{D}_2(\cdot|\xi_{[1]})}\Bigg[
        \sum_{i,j,b} Q_2 + \cdots
\nonumber\\
&\qquad\qquad \cdots
    + \mathbb{E}_{\xi_{T} \sim \mathcal{D}_{T}(\cdot|\xi_{[T-1]})}\Bigg[
        \sum_{i,j,b} Q_T \;+\; \Phi\!\left(x_{ijbT}^{\mathrm{r}},
        x_{ijbT}^{\mathrm{p}}, x_{iibT}^{\mathrm{c}}, 
        x_{iibT}^{\mathrm{d}}\right)
    \Bigg]\cdots\Bigg]\Bigg],
\end{align}
where we use the shorthand 
$Q_t \triangleq Q(x_{ijbt}^{\mathrm{r}}, x_{ijbt}^{\mathrm{p}}, 
x_{iibt}^{\mathrm{c}}, x_{iibt}^{\mathrm{d}})$ for brevity.

The stage cost $Q_t$ is defined as
\begin{equation}
\label{eq:stagecost}
Q\!\left(x_{ijbt}^{\mathrm{r}}, x_{ijbt}^{\mathrm{p}}, 
         x_{iibt}^{\mathrm{c}}, x_{iibt}^{\mathrm{d}}\right)
= c_{ijb}^{\mathrm{r}}\,x_{ijbt}^{\mathrm{r}}
- c_{ijb}^{\mathrm{p}}\,x_{ijbt}^{\mathrm{p}}
+ c_{ibt}^{\mathrm{c}}\,x_{iibt}^{\mathrm{c}}
- c_{ibt}^{\mathrm{d}}\,x_{iibt}^{\mathrm{d}},
\end{equation}
where the cost coefficients $c_{ijb}^{\mathrm{r}}, c_{ijb}^{\mathrm{p}}, 
c_{ibt}^{\mathrm{c}}, c_{ibt}^{\mathrm{d}}$ are defined in 
Table~\ref{tab:optivar}.

The terminal cost $\Phi$ penalizes vehicles whose state of charge (SoC) 
is below the maximum level $b_{\max}$ at the end of the horizon:
\begin{equation}
\label{eq:terminalcost}
\Phi = \min\!\bigl(q_{\max},\;\max\!\bigl((b_{\max}-b)\,a,\;0\bigr)\bigr)
\left(
  \sum_{i,j,b} x_{ijbT}^{\mathrm{r}} + x_{ijbT}^{\mathrm{p}}
  + \sum_{i,b}  x_{iibT}^{\mathrm{c}} + x_{iibT}^{\mathrm{d}}
\right).
\end{equation}
This term accounts for continued operation of the EAMoD system beyond 
the optimization horizon. The constant $q_{\max}$ bounds the penalty, 
and the parameter $a$ converts the battery shortfall $(b_{\max}-b)$ 
into a monetary cost per unit of missing energy, proportional to 
anticipated future service requirements and electricity prices.

The nested expectation structure in \eqref{eq:smpcobjective} reflects 
the sequential nature of the problem: the cost at stage $t$ is incurred 
after observing $\xi_{[t]}$ but before $\xi_{[t+1]}$ is revealed, so 
each inner expectation conditions on the history available at that 
stage. This is the standard formulation of multi-stage stochastic 
programming \cite{shapiro2011analysis}.

A key challenge in EAMoD optimization is the mismatch between short-term trip dynamics (minutes) and longer-term charging economics (hours to a full day). We address this by leveraging day-ahead electricity prices, which are known in advance, to dynamically adjust the terminal cost parameter $a$ at each receding horizon iteration. Let $p_t$ denote the electricity price at time $t$ and let $\mathcal{T}^+ = \{T+1,\ldots,T+H\}$ denote the lookahead window beyond the prediction horizon, where $H$ is the lookahead length. The adaptive terminal cost parameter is computed as:
\begin{equation}\label{eq:adaptiveterminal}
a_t = a_{\text{base}} \cdot \frac{\bar{p}^+}{\max(p_t,\, p_{\min})}, \qquad \text{where} \quad \bar{p}^+ = \frac{1}{H}\sum_{\tau \in \mathcal{T}^+} p_\tau.
\end{equation}
Here, $a_{\text{base}}$ is the baseline terminal cost parameter, $p_t$ is the current electricity price, $\bar{p}^+$ is the average electricity price over the lookahead window, and  $p_{\min} > 0$ is a minimum price threshold that prevents unbounded terminal cost scaling during periods of very low or zero electricity prices (e.g., due to excess solar generation). When future prices exceed current prices ($\bar{p}^+ > p_t$), the ratio increases $a_t$, incentivizing immediate charging. Conversely, when future prices are lower ($\bar{p}^+ < p_t$), the reduced $a_t$ encourages deferring charging to take advantage of cheaper electricity. This adaptive terminal cost effectively encodes day-ahead price information into the short-horizon optimization, enabling the SMPC to coordinate charging schedules over daily time scales without explicitly extending the prediction horizon.

\begin{table}
\renewcommand{\arraystretch}{1.2}
\caption{Cost and profit components used in the optimization model.}
\label{tab:optivar}
\begin{tabularx}{\textwidth}{l|X}
\hline
\textbf{Symbol} & \textbf{Description} \\ \hline
$c_{ijb}^{\text{r}}$   & The cost of rebalancing a vehicle from station $i$ to station $j$ with SoC $b$ at time $t$, representing the expected energy consumption cost. \\
$c_{ibt}^{\text{c}}$   & The cost of fast charging a vehicle at station $i$ at time $t$, where the electricity price may vary depending on the time of day. \\
$c_{ijb}^{\text{p}}$   & The profit from transporting customers from station $i$ to station $j$ with SoC $b$ at time $t$. This profit is calculated based on the distance, with a constant cost per distance throughout the day. \\
$c_{ibt}^{\text{d}}$ & The profit from V2G services provided by a vehicle at station $i$ at time $t$. The profitability is influenced by the varying electricity prices throughout the day. \\ \hline
\end{tabularx}
\end{table}

\subsection{The Multi-Stage SMPC Problem}\label{sec:smpc}

The EAMoD control problem for rebalancing and charging under uncertainty can now be formulated using the model and objective function introduced above. The overall goal is to improve both service quality and operational efficiency. This is achieved by strategically rebalancing vehicles to match spatial demand patterns and by scheduling charging and V2G operations to maintain adequate battery SoC. 

Uncertainty in the future evolution of the random variables is addressed using a combined scenario-based and robust approach. Travel demand and charger availability are modeled by a scenario tree that approximates the expected value, with each realization indexed by the superscript $s$. Uncertainties in energy consumption and travel time are handled through the chance-constrained parameterization of the model parameters $\Delta b_{ijt}$, $\Delta t_{jit}$, and $b_{ij}^{\min}$ described in Section~\ref{sec:flowdyn}, which ensures operational feasibility with high probability. The resulting optimization problem is updated sequentially as new information about the random variables becomes available. The multi-stage SMPC problem is given by
\begin{subequations}\label{eq:evmpc}
\begin{align}
        &\min_{x_{ijbt}^{\text{r},s},\, x_{ijbt}^{\text{p},s},\, x_{iibt}^{\text{c},s},\, x_{iibt}^{\text{d},s}} 
        J_N(x_{ijbt}^{\text{r},s}, x_{ijbt}^{\text{p},s}, x_{iibt}^{\text{c},s}, x_{iibt}^{\text{d},s})\\
        & \text{subject to} \notag \\
        & \eqref{eq:flowcons},\ \eqref{eq:im},\ \eqref{eq:charger}, \\
        & x_{ijbt}^{\text{r},s} = x_{ijbt}^{\text{r},s'} ,\quad 
   x_{ijbt}^{\text{p},s} = x_{ijbt}^{\text{p},s'} ,\quad 
   x_{iibt}^{\text{c},s} = x_{iibt}^{\text{c},s'} ,\quad 
   x_{iibt}^{\text{d},s} = x_{iibt}^{\text{d},s'} \qquad \forall s, s' \in S_t:\ (\xi^s_0,...,\xi^s_{t-1})=(\xi^{s'}_0,...,\xi^{s'}_{t-1}), \label{eq:non-antici1} \\
        & x_{ijbt}^{\text{p}},\ x_{iibt}^{\text{c}},\ x_{iibt}^{\text{d}},\ s_{ijt} \in \mathbb{N} \qquad \forall i,j \in \mathcal{N},\ t \in \mathcal{T},\ b \in \mathcal{B},\\
        & x_{ijbt}^{\text{r}} \in \mathbb{R}^+ \qquad \forall i,j \in \mathcal{N},\ t \in \mathcal{T},\ b \in \mathcal{B}.
\end{align}
\end{subequations}
Constraint \eqref{eq:non-antici1} enforces non-anticipativity: at each stage, decisions must be identical across all scenarios that share the same history. Note that the uncertainties in energy consumption and travel time do not appear as explicit constraints in the optimization. Instead, they are handled through the chance-constrained parameterization of the model described in Section~\ref{sec:flowdyn}: the conservative travel time estimates $\hat{\tau}_{jit}$ from~\eqref{eq:cctime_model} enter the flow conservation constraint~\eqref{eq:flowcons}, and the minimum SoC threshold $b_{ij}^{\min}$ from~\eqref{eq:bmin_def} restricts vehicle dispatch in the imbalance constraint~\eqref{eq:im}. These parameters are computed offline at each receding horizon iteration before the optimization is solved, effectively embedding the robust treatment of energy and travel time uncertainty into the model structure itself.

The problem is nearly totally unimodular (TU). In general, the LP relaxation of a TU mixed-integer linear program (MILP) yields integer solutions \citep{Hoffman2010}. In our case, however, two constraints break TU. First, Eq.~\eqref{eq:im} enforces integrality on the \emph{sum} of passenger-serving variables $x_{ijbt}^{\text{p}}$ rather than on each variable individually. Second, Eq.~\eqref{eq:charger} couples charging and discharging decisions across vehicles in a way that also violates TU. Consequently, explicit integrality must be imposed on $x_{ijbt}^{\text{p}}, x_{iibt}^{\text{c}},$ and $x_{iibt}^{\text{d}}$. Only the rebalancing variables $x_{ijbt}^{\text{r}}$ can be treated as positive continuous flows. The overall optimization problem therefore remains a MILP but cannot exploit full TU structure. A significant computational challenge in solving Eq.~\eqref{eq:evmpc} arises from the exponential growth in the number of scenarios considered at each stage, as also noted by \citep{richards2005mixed}. To address this, the next section introduces an integrated scenario prediction and reduction method, together with a decomposition technique.
The resulting extensive-form MILP grows exponentially with the number  of stages and scenarios, rendering direct solution impractical for real-time MPC applications.

\section{Scenario Tree Generation and Reduction} \label{section:scentree}
To optimize fleet rebalancing and charging under uncertainty, we use a SMPC framework. A key challenge is the uncertain evolution of future travel demand and charger availability, which prevents a closed-form evaluation of the expected cost. We address this by representing uncertainty with a scenario tree that captures both multiple realizations and temporal dependence. This section outlines the construction of the scenario tree, the Bayesian Neural Network (BNN)-based forecasting methodology, the scenario reduction procedure, and the integration of the reduced tree into the SMPC formulation.

\subsection{Scenario Tree Construction}
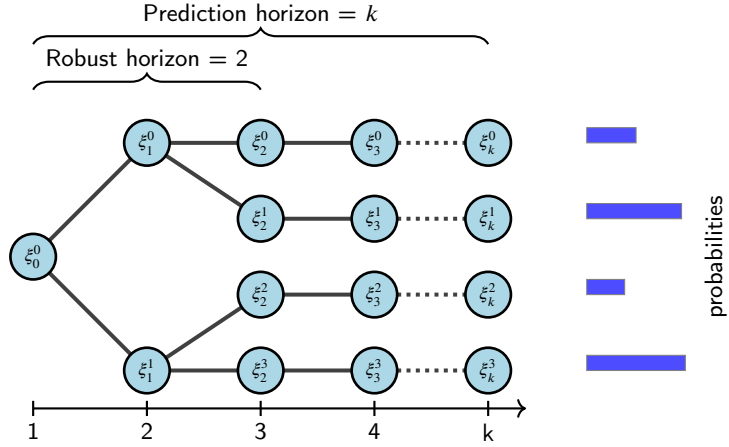
\begin{figure}
    \centering 
    \begin{tikzpicture}
       \begin{scope}[] 

        \Vertex[x=0 ,y=0,label=$\xi_0^0$]{A}
        \Vertex[x=1.5 ,y=1.5,label=$\xi_1^0$]{B1}
        \Vertex[x=1.5 ,y=-1.5,label=$\xi_1^1$]{B2}
        \Vertex[x=3 ,y=1.5,label=$\xi_2^0$]{C1}
        \Vertex[x=3 ,y=0.5,label=$\xi_2^1$]{C2}
        \Vertex[x=3 ,y=-0.5,label=$\xi_2^2$]{C3}
        \Vertex[x=3 ,y=-1.5,label=$\xi_2^3$]{C4}
        \Vertex[x=4.5 ,y=1.5,label=$\xi_3^0$,]{D1}
        \Vertex[x=4.5 ,y=0.5,label=$\xi_3^1$]{D2}
        \Vertex[x=4.5 ,y=-0.5,label=$\xi_3^2$]{D3}
        \Vertex[x=4.5 ,y=-1.5,label=$\xi_3^3$]{D4}
        
        \Vertex[x=6 ,y=1.5,label=$\xi_k^0$]{E1}
        \Vertex[x=6 ,y=0.5,label=$\xi_k^1$]{E2}
        \Vertex[x=6 ,y=-0.5,label=$\xi_k^2$]{E3}
        \Vertex[x=6 ,y=-1.5,label=$\xi_k^3$]{E4}
        
        \Edge(A)(B1)
        \Edge(A)(B2)
        \Edge[](B1)(C1)
        \Edge[](B1)(C2)
        \Edge[](B2)(C3)
        \Edge[](B2)(C4)
        \Edge[](D1)(C1)
        \Edge[](D2)(C2)
        \Edge[](D3)(C3)
        \Edge[](D4)(C4)
        \Edge[style={dotted}](D1)(E1)
        \Edge[style={dotted},](D2)(E2)
        \Edge[style={dotted},](D3)(E3)
        \Edge[style={dotted},](D4)(E4)
      \end{scope}

      \draw[->, thick,black] (0, -2) -- (6.5, -2) node[anchor=north west] {};
      \draw[thick,black] (0, -2.1) -- (0, -1.9) node[anchor=north, yshift=-5pt] {1};
      \draw[thick,black] (1.5, -2.1) -- (1.5, -1.9) node[anchor=north, yshift=-5pt] {2};
      \draw[thick,black] (3, -2.1) -- (3, -1.9) node[anchor=north, yshift=-5pt] {3};
      \draw[thick,black] (4.5, -2.1) -- (4.5, -1.9) node[anchor=north, yshift=-5pt] {4};
      \draw[thick,black] (6, -2.1) -- (6, -1.9) node[anchor=north, yshift=-5pt] {k};
    \draw [decorate,decoration={brace,amplitude=6pt},thick,color=black]
    (0,2.2) -- (3,2.2) node [midway, yshift=12pt] {Robust horizon = 2};
    
    \draw [decorate,decoration={brace,amplitude=6pt},thick,color=black]
    (0,2.8) -- (6,2.8) node [midway, yshift=12pt] {Prediction horizon = $k$};
    \draw[fill=blue!70] (7.3,1.5) rectangle ++(0.65,0.2); 
    \draw[fill=blue!70] (7.3,0.5) rectangle ++(1.25,0.2); 
    \draw[fill=blue!70] (7.3,-0.5) rectangle ++(0.50,0.2); 
    \draw[fill=blue!70] (7.3,-1.5) rectangle ++(1.30,0.2); 
    
    \node[rotate=90,color=black] at (9.0, 0) {probabilities};
    
    \end{tikzpicture}
    \caption{Scenario tree representation of the random variables, incorporating both the robust horizon and prediction horizon. Each scenario is defined as $\xi^s_t=(\lambda^s_{ijt},k_{i,t}^{c,s})$.}
    \label{fig:scentreerob}
\end{figure}

The true distributions \(\mathcal{D}_{\lambda}^*\) and \(\mathcal{D}_{\text{k}}^*\) for travel demand and charger availability are unknown, so the expectation in \eqref{eq:smpcobjective} cannot be computed directly. To approximate this, we represent uncertainty using a scenario tree, which encodes multiple realizations of the random variables and their evolution over time (see Fig. \ref{fig:scentreerob}). At each time step \(t\), a scenario \(s \in \{1,2,\dots,S\}\) is defined as \(\xi^s_t=[\lambda_{ijt}^s, k_{it}^{c,s}]\), where \(\lambda_{ijt}^s\) is the predicted travel demand and \(k_{it}^{c,s}\) the available chargers in scenario \(s\). The cost function is then reformulated using the Sample Average Approximation (SAA) approach:

\begin{align} \label{eq:smpcobjective_saa}
\hat{J}_N(x_{ijbt}^{\text{r},s}, x_{ijbt}^{\text{p},s}, x_{iibt}^{\text{c},s}, x_{iibt}^{\text{d},s}) = & \sum_{i=1}^N\sum_{j=1}^N\sum_{b=0}^{1} Q(x_{ijb0}^{\text{r}}, x_{ijb0}^{\text{p}}, x_{iib0}^{\text{c}}, x_{iib0}^{\text{d}}) \\
& + \sum_{t=1}^{T-1}\sum_{s=1}^{S_t} p_s \sum_{i=1}^N\sum_{j=1}^N\sum_{b=0}^{1} Q(x_{ijbt}^{\text{r},s}, x_{ijbt}^{\text{p},s}, x_{iibt}^{\text{c},s}, x_{iibt}^{\text{d},s}) \nonumber \\
& + \sum_{s=1}^{S_T} p_s \sum_{i=1}^N\sum_{j=1}^N\sum_{b=0}^{1} C^{\text{term}}(b) \cdot \left(x_{ijbT}^{\text{r},s} + x_{ijbT}^{\text{p},s} + x_{iibT}^{\text{c},s} + x_{iibT}^{\text{d},s}\right), \nonumber
\end{align}
where \(p_s\) denotes the probability of scenario \(s\), and \(\sum_{s=1}^S p_s = 1\). The function \(\hat{J}_N(\cdot)\) approximates the original objective in Eq.~\eqref{eq:smpcobjective}, where the expected value is replaced with a scenario tree. To avoid exponential growth of the scenario tree with respect to the prediction horizon, we introduce the concept of a \textit{robust horizon}, beyond which uncertainty is assumed to remain constant. This idea, proposed in~\citep{lucia2013multi}, allows branching only up to a certain stage, making the tree more tractable (see Fig. \ref{fig:scentreerob}). Additionally, to further limit unnecessary tree growth when the time discretization is fine, scenarios are assumed to remain constant over short time intervals, effectively reducing the number of unique branches in closely spaced time steps.

\subsection{Forecasting Travel Demand and Charger Availability}
Given historical spatio-temporal data on travel demand and charger availability, our goal is to predict future values, quantify uncertainty, and model correlations across time and space. Input features \(x\) include time, location, and contextual variables such as weather or day of the week. Travel demand is recorded as the number of trips between discretized spatial zones over fixed time intervals, resulting in matrices \(\lambda_{ijt} \in \mathbb{Z}^{N \times N}\). Charger availability is given as vectors \(k_{it}^{c} \in \mathbb{Z}^N\). Because both outputs are non-negative counts, the predictive model must produce integer-valued outputs and capture predictive uncertainty.

We use a Bayesian Neural Network (BNN) to model these forecasts. Unlike standard neural networks, BNNs maintain distributions over their weights \(\theta\), allowing for explicit modeling of epistemic uncertainty. Given input features \(x\), the neural network defines a parametric function \(f(x; \theta)\) that produces an output \(\hat{y}\). The Bayesian approach introduces a posterior distribution over the parameters:

\begin{equation}
    p(\theta | \mathcal{D}) \propto p(\mathcal{D} | \theta) p(\theta),
\end{equation}

where \(p(\theta)\) is the prior over weights and \(p(\mathcal{D} | \theta)\) is the likelihood of the observed data given parameters \(\theta\). Prediction for a new input \(x\) involves marginalizing over this posterior:
\begin{equation}
    p(\hat{y} | x, \mathcal{D}) = \int p(\hat{y} | x, \theta) p(\theta | \mathcal{D}) d\theta.
\end{equation}
This integral is intractable in general, so approximate inference techniques such as Variational Inference (VI) or Markov Chain Monte Carlo (MCMC) are used. In our work, we adopt the BNF model proposed in~\citep{saad2024scalable}, which is specifically designed for spatio-temporal prediction with multiple correlated outputs. It models the predictive distribution using the Negative Binomial (NB) distribution, which is well-suited for overdispersed count data:
\[
p(\hat{y}(x) | \mathcal{D}) \sim \mathrm{NB}(\hat{\mu}, \hat{\sigma}^2),
\]
where \(\hat{\mu}\) and \(\hat{\sigma}^2\) are the mean and variance predicted by the model. These predictive distributions form the basis for scenario generation in the next step of our stochastic optimization framework.
An important advantage of the Bayesian formulation is that the model can be updated sequentially as new data arrive. In the variational inference setting, this is achieved by using the previously learned posterior as the prior when incorporating a new dataset 
\( \mathcal{D}_{t} \). This enables efficient online learning without retraining from scratch, while preserving past information. Such posterior-to-prior updates allow the BNN to adapt to changing travel demand patterns or charger availability in real time, which is crucial for EAMoD operations. These predictive distributions, updated over time, form the basis for scenario generation in the next step of our stochastic optimization framework.

 The following summarizes the key hyperparameters of the BNF model used in our experiments. A daily seasonality period with 7 harmonics captures intra-day demand patterns such as morning and evening peaks. The feature set consists of five columns: the time interval index and the spatial coordinates of both pickup $(x, y)$ and dropoff $(x_d, y_d)$ locations. To capture joint spatio-temporal effects, all $\binom{5}{2} = 10$ pairwise feature interactions are included as explicit second-order cross-product inputs to the network. These interactions fall into three categories: (i)~ spatiotemporal coupling (e.g., time $\times$ pickup longitude), which allows demand at different locations to vary independently over the day; (ii)~origin--destination flow structure (e.g., pickup longitude $\times$ dropoff longitude), which captures directional travel patterns between regions; and (iii)~neighborhood identification (e.g., pickup $x \times$ pickup $y$), which enables the model to distinguish specific geographic areas. By providing these cross-products explicitly, the model can learn multiplicative relationships efficiently even with a compact architecture (width 64, depth 2), rather than relying solely on the hidden layers to discover them.

\subsection{Scenario Tree Reduction}
From the BNF model, we generate scenario samples of the form:
\[
\xi^s = (\xi_1^s, \xi_2^s, \dots, \xi_T^s), \quad s = 1, \dots, S,
\]
where each \(\xi_t^s\) consists of sampled travel demand and charger availability values for scenario \(s\) at time \(t\). This collection of sampled sequences is referred to as a \textit{scenario fan}. To ensure tractability, we reduce the number of scenarios using clustering techniques. One common approach is to treat the problem as a two-stage stochastic program, but this requires a large number of samples. Instead, we construct a reduced scenario tree that preserves the structure of a multi-stage stochastic process. The tree is built iteratively from the root. At each stage, we sample future values conditionally using the predictive model:
\[
\xi^s_{(t+1)} \sim P(\hat{y} \mid \mathcal{D}, \hat{y}(x(t))).
\]
We then apply soft \(k\)-means clustering to reduce the number of scenarios, using the cluster centroids as representatives for each scenario. This process is repeated for each time stage. The number of clusters \(K_t\) at each step is specified by the user in order to balance tractability and representativeness of the uncertainty. 

Alternative reduction approaches based on probability metrics, such as the Wasserstein distance or the Nested distance, are not adopted here due to their high computational cost and limited scalability in multi-stage settings. Instead, our method follows the general idea of clustering-based reductions that have been proposed in previous works, where scenarios are grouped and replaced by representative centroids. This allows us to construct reduced scenario trees that remain computationally feasible while still capturing the main patterns of uncertainty. The full reduction algorithm is given in Algorithm~\ref{alg:scenario_tree}.


\begin{algorithm}
\caption{Scenario Tree Reduction}
\label{alg:scenario_tree}
\begin{algorithmic}[1]
    \State \textbf{Input:} Initial state \(\xi_0\), dataset \(\mathcal{D}\), prediction model \(P(\cdot \mid \mathcal{D})\), time horizon \(T\), number of scenarios per stage \(K_t\)
    \State \textbf{Output:} Reduced scenario tree
    \State Initialize root node with \(\xi_0\)
    \For{\(t = 0\) to \(T - 1\)}
        \State Sample \(M\) scenarios: \(\xi^s(t+1) \sim P(\hat{y}(x) \mid \mathcal{D}, \hat{y}(x(t)))\)
        \State Apply soft \(k\)-means to obtain \(K_t\) representative scenarios (with \(K_t\) specified by the user)
        \State Set cluster centroids 
        \State Assign transition probabilities from clustering weights
    \EndFor
    \State \Return Constructed scenario tree
\end{algorithmic}
\end{algorithm}

\section{Nested Benders Decomposition} \label{sec:NB}
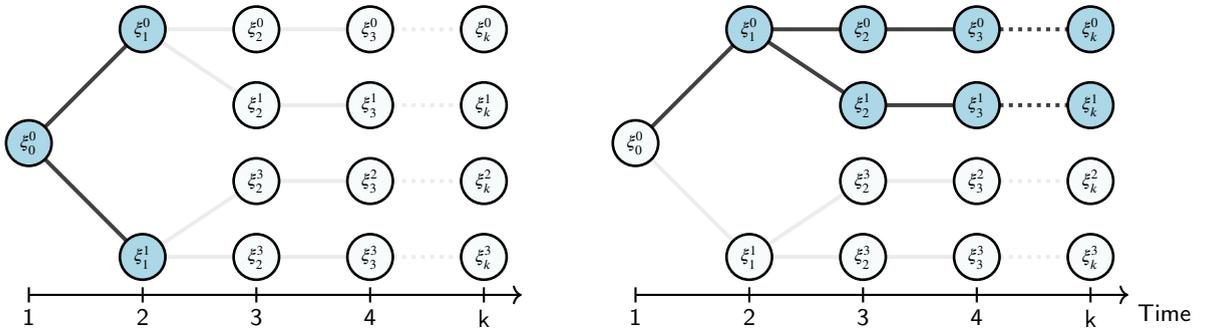
\begin{figure}
    \centering
\begin{tikzpicture}
   \begin{scope}[shift={(0,0)}] 

    \Vertex[x=0 ,y=0,label=$\xi_0^0$]{A}
    \Vertex[x=1.5 ,y=1.5,label=$\xi_1^0$]{B1}
    \Vertex[x=1.5 ,y=-1.5,label=$\xi_1^1$]{B2}
    \Vertex[x=3 ,y=1.5,label=$\xi_2^0$,opacity=0.1]{C1}
    \Vertex[x=3 ,y=0.5,label=$\xi_2^1$,opacity=0.1]{C2}
    \Vertex[x=3 ,y=-0.5,label=$\xi_2^3$,opacity=0.1]{C3}
    \Vertex[x=3 ,y=-1.5,label=$\xi_2^3$,opacity=0.1]{C4}
    \Vertex[x=4.5 ,y=1.5,label=$\xi_3^0$,opacity=0.1]{D1}
    \Vertex[x=4.5 ,y=0.5,label=$\xi_3^1$,opacity=0.1]{D2}
    \Vertex[x=4.5 ,y=-0.5,label=$\xi_3^2$,opacity=0.1]{D3}
    \Vertex[x=4.5 ,y=-1.5,label=$\xi_3^3$,opacity=0.1]{D4}
    \Vertex[x=6 ,y=1.5,label=$\xi_k^0$,opacity=0.1]{E1}
    \Vertex[x=6 ,y=0.5,label=$\xi_k^1$,opacity=0.1]{E2}
    \Vertex[x=6 ,y=-0.5,label=$\xi_k^2$,opacity=0.1]{E3}
    \Vertex[x=6 ,y=-1.5,label=$\xi_k^3$,opacity=0.1]{E4}

    \Edge(A)(B1)
    \Edge(A)(B2)
    \Edge[opacity=0.1](B1)(C1)
    \Edge[opacity=0.1](B1)(C2)
    \Edge[opacity=0.1](B2)(C3)
    \Edge[opacity=0.1](B2)(C4)
    \Edge[opacity=0.1](D1)(C1)
    \Edge[opacity=0.1](D2)(C2)
    \Edge[opacity=0.1](D3)(C3)
    \Edge[opacity=0.1](D4)(C4)
    \Edge[style={dotted},opacity=0.1](D1)(E1)
    \Edge[style={dotted},opacity=0.1](D2)(E2)
    \Edge[style={dotted},opacity=0.1](D3)(E3)
    \Edge[style={dotted},opacity=0.1](D4)(E4)
  \end{scope}

 \begin{scope}[shift={(8,0)}] 

    \Vertex[x=0 ,y=0,label=$\xi_0^0$,opacity=0.1]{A}
    \Vertex[x=1.5 ,y=1.5,label=$\xi_1^0$]{B1}
    \Vertex[x=1.5 ,y=-1.5,label=$\xi_1^1$,opacity=0.1]{B2}
    \Vertex[x=3 ,y=1.5,label=$\xi_2^0$]{C1}
    \Vertex[x=3 ,y=0.5,label=$\xi_2^1$]{C2}
    \Vertex[x=3 ,y=-0.5,label=$\xi_2^3$,opacity=0.1]{C3}
    \Vertex[x=3 ,y=-1.5,label=$\xi_2^3$,opacity=0.1]{C4}
    \Vertex[x=4.5 ,y=1.5,label=$\xi_3^0$]{D1}
    \Vertex[x=4.5 ,y=0.5,label=$\xi_3^1$]{D2}
    \Vertex[x=4.5 ,y=-0.5,label=$\xi_3^2$,opacity=0.1]{D3}
    \Vertex[x=4.5 ,y=-1.5,label=$\xi_3^3$,opacity=0.1]{D4}
    \Vertex[x=6 ,y=1.5,label=$\xi_k^0$]{E1}
    \Vertex[x=6 ,y=0.5,label=$\xi_k^1$]{E2}
    \Vertex[x=6 ,y=-0.5,label=$\xi_k^2$,opacity=0.1]{E3}
    \Vertex[x=6 ,y=-1.5,label=$\xi_k^3$,opacity=0.1]{E4}

    \Edge([opacity=0.1]A)(B1)
    \Edge[opacity=0.1](A)(B2)
    \Edge(B1)(C1)
    \Edge(B1)(C2)
    \Edge[opacity=0.1](B2)(C3)
    \Edge[opacity=0.1](B2)(C4)
    \Edge(D1)(C1)
    \Edge(D2)(C2)
    \Edge[opacity=0.1](D3)(C3)
    \Edge[opacity=0.1](D4)(C4)
    \Edge[style={dotted}](D1)(E1)
    \Edge[style={dotted}](D2)(E2)
    \Edge[style={dotted},opacity=0.1](D3)(E3)
    \Edge[style={dotted},opacity=0.1](D4)(E4)
  \end{scope}
  
      \draw[->, thick,black] (0, -2) -- (6.5, -2) node[anchor=north west] [text=black] {};
      
      \draw[thick,black] (0, -2.1) -- (0, -1.9) node[anchor=north, yshift=-5pt] {1};
      \draw[thick,black] (1.5, -2.1) -- (1.5, -1.9) node[anchor=north, yshift=-5pt] {2};
      \draw[thick,black] (3, -2.1) -- (3, -1.9) node[anchor=north, yshift=-5pt] {3};
      \draw[thick,black] (4.5, -2.1) -- (4.5, -1.9) node[anchor=north, yshift=-5pt] {4};
      \draw[thick,black] (6, -2.1) -- (6, -1.9) node[anchor=north, yshift=-5pt] {k};
        
      \draw[->, thick,black] (8, -2) -- (14.5, -2) node[anchor=north west] [text=black] {Time};
      
      \draw[thick,black] (8, -2.1) -- (8, -1.9) node[anchor=north, yshift=-5pt] {1};
      \draw[thick,black] (9.5, -2.1) -- (9.5, -1.9) node[anchor=north, yshift=-5pt] {2};
      \draw[thick,black] (11, -2.1) -- (11, -1.9) node[anchor=north, yshift=-5pt] {3};
      \draw[thick,black] (12.5, -2.1) -- (12.5, -1.9) node[anchor=north, yshift=-5pt] {4};
      \draw[thick,black] (14,-2.1) -- (14, -1.9) node[anchor=north, yshift=-5pt] {k};
\end{tikzpicture}
    \caption{An example of solving a multi-stage SMPC problem by decomposing it into a nested series of two-stage SMPC problems. The figure on the left illustrates the first and second stages of the SMPC problem, while the figure on the right shows a portion of the second and third stages. Both sections can be resolved using Benders Decomposition.}
    \label{fig:nested-benders-three}
\end{figure}

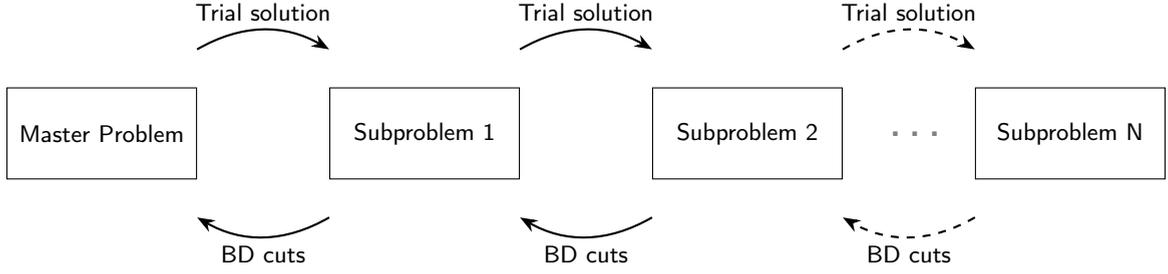
\begin{figure}
    \centering
\begin{tikzpicture}[
  box/.style={
    draw=black,
    text=black,
    minimum width=2.5cm,
    minimum height=1.2cm,
    align=center
  },
  arrow/.style={draw=black, thick, -{Stealth}},
  labelstyle/.style={font=\small, text=black}
]

\node[box] (box1) at (0,0) {Master Problem};
\node[box] (box2) at ($(box1.east)+(3,0)$) {Subproblem 1};
\node[box] (box3) at ($(box2.east)+(3,0)$) {Subproblem 2};
\node[draw=none] (dots) at ($(box3.east)+(1,0)$) {\Large\bfseries \dots};
\node[box] (boxN) at ($(dots)+(2,0)$) {Subproblem N};

\draw[arrow] ([yshift=5mm]box1.north east) to [bend left=30] node[labelstyle, above, midway] {Trial solution} ([yshift=5mm] box2.north west);
\draw[arrow] ([yshift=-5mm]box2.south west) to [bend left=30] node[labelstyle, below, midway] {BD cuts} ([yshift=-5mm]box1.south east);

\draw[arrow] ([yshift=5mm]box2.north east) to [bend left=30] node[labelstyle, above, midway] {Trial solution} ([yshift=5mm]box3.north west);
\draw[arrow] ([yshift=-5mm]box3.south west) to [bend left=30] node[labelstyle, below, midway] {BD cuts} ([yshift=-5mm]box2.south east);

\draw[arrow, dashed] ([yshift=5mm]box3.north east) to [bend left=30] node[labelstyle, above, midway] {Trial solution} ([yshift=5mm]boxN.north west);
\draw[arrow, dashed] ([yshift=-5mm]boxN.south west) to [bend left=30] node[labelstyle, below, midway] {BD cuts} ([yshift=-5mm]box3.south east);

\end{tikzpicture}
    \caption{Illustration of nested Benders decomposition. The master problem proposes trial solutions which are passed sequentially to subproblems at different stages. Each subproblem generates Benders cuts based on its solution and returns them to the previous stage, enabling iterative refinement across the scenario tree}
    \label{fig:l-shaped method}
\end{figure}

Directly solving the multi-stage stochastic problem in Eq.~\eqref{eq:evmpc} can be computationally expensive, and even intractable, due to the exponential growth of scenarios over the planning horizon. To address this challenge, we employ Nested Benders Decomposition (NBD), an extension of classical Benders decomposition tailored for multi-stage stochastic programs. The key idea is to decompose the original problem into a hierarchy of two-stage problems, one for each stage and scenario, while iteratively exchanging information between them (see Figs. \ref{fig:nested-benders-three} and \ref{fig:l-shaped method}). This approach reduces the computational burden while preserving anticipativity across the scenario tree.

In the NBD framework, the problem is reformulated as a sequence of nested two-stage subproblems, each associated with a specific stage and scenario (Fig.~\ref{fig:nested-benders-three}). A master problem (MP) coordinates the solution process by integrating Benders cuts derived from the duals of the subproblems (SP) (Fig.~\ref{fig:l-shaped method}). These cuts iteratively refine the feasible region of the MP, guiding the solution process toward global optimality while reducing computational effort.

The NBD procedure is executed in a forward–backward iterative scheme, commonly referred to as fast-forward–fast-back sequencing \citep{murphy2013benders}. During the forward pass, the first-stage problem and subsequent stages, up to one stage before the robust horizon, are treated as master problems. In the backward pass, only the first stage remains a master problem, while all subsequent stages are treated as subproblems. To compute valid duals, integer variables in these subproblems must be relaxed to continuous ones.

The iterative process terminates once convergence is achieved, defined as the gap between the subproblem costs and their estimated values in the MP falling below a specified tolerance, or when a maximum number of iterations is reached. Because subproblem integer variables are relaxed, the resulting solutions may not be integer-feasible. This relaxation introduces a \emph{duality gap} between the relaxed subproblem objective and the true integer-optimal objective, which weakens the tightness of the Benders cuts. Techniques such as Gomory cuts \citep{gade2014decomposition} and Lagrangian cuts \citep{chen2022generating} can yield tighter cuts, but they often significantly slow convergence and are therefore not adopted in our implementation.    

\newpage
Within the NBD framework, multiple MPs are defined, and as the decision process progresses down the scenario tree, SPs are recursively promoted to MPs. To formalize this structure, we present general formulations for both MPs and SPs. Based on the SMPC formulation in Eq.~\ref{eq:evmpc} and the reformulated objective function in Eq.~\ref{eq:smpcobjective_saa}, the first-stage master problem can be expressed as follows:

\begin{subequations}\label{eq:mp1evmpc}
\begin{align}
        &\minimize_{x_{ijbt}^{\text{r},s}, x_{ijbt}^{\text{p},s}, x_{iibt}^{\text{c},s}, x_{iibt}^{\text{d},s}} 
        J_N(x_{ijbt}^{\text{r},s}, x_{ijbt}^{\text{p},s}, x_{iibt}^{\text{c},s}, x_{iibt}^{\text{d},s}) + \sum_{s=1}^S p_s \alpha_s\\
        & \text{subject to} \notag \\
        & \eqref{eq:flowcons},\ \eqref{eq:im},\ \eqref{eq:charger},\ \eqref{eq:non-antici1} \\
        & \alpha_s \geq Q^{s} + \sum_{ijb} \pi_{ijbt}^{\text{r},s}(x_{ijbt}^{\text{r},s}-\bar{x}_{ijbt}^{\text{r},s}))+\pi_{ijbt}^{\text{p},s}(x_{ijbt}^{\text{p},s}-\bar{x}_{ijbt}^{\text{p},s}))+\pi_{iibt}^{\text{c},s}(x_{iibt}^{\text{c},s}-\bar{x}_{iibt}^{\text{c},s}))+\pi_{iibt}^{\text{d},s}(x_{iibt}^{\text{d},s}-\bar{x}_{iibt}^{\text{d},s}))\label{eq:masterBDcut} \\
        & \qquad \forall s \in S_{t+1} \\
        &x_{ijbt}^{\text{p}} \in \mathbb{N} \qquad \forall i,j \in N,t \in \mathcal{T}, \forall b \in \mathcal{B}.\\
        &x_{ijbt}^{\text{r}},x_{iibt}^{\text{c}},x_{iibt}^{\text{d}} \in \mathbb{R}^+ \qquad \forall i,j \in \mathcal{N},\forall t \in \mathcal{T}, \forall b \in \mathcal{B},
\end{align}
\end{subequations}
where $p_s$ is the probability of each scenario, $S_{t}$ is the number of scenarios in the stage, and $\alpha_s$ is the approximated lower bound of the cost of the subproblem. The MP inherits all structural constraints from the SMPC formulation \eqref{eq:evmpc}, including flow conservation, imbalance dynamics, charger capacity, and non-anticipativity. The chance constraints on energy consumption and travel time are embedded in the model parameters (see Section~\ref{sec:flowdyn}) and thus implicitly enforced through~\eqref{eq:flowcons} and~\eqref{eq:im}. These constraints are augmented by the Benders cuts in constraint~\eqref{eq:masterBDcut}, which approximate the future cost. Constraint \eqref{eq:masterBDcut} represents the Benders cuts from the subproblems at the next time step, and $S_{t+1}$ denotes the set of all subproblems at time $t+1$. This version of the Benders cuts is referred to as the compact form. In this constraint, $Q^s$ represents the cost of each next-stage subproblem, and $\pi$ denotes the dual variables associated with the corresponding decision variables in the MP. The BD cuts generated in these MPs will always be optimality cuts due to the flow conservation constraints in Eq.~\eqref{eq:flowcons}. As a result, feasibility cuts are not required. In this implementation, we use multi-cuts, where one Benders cut is added to the master problem for each subproblem. Alternatively, a single-cut approach can be used, where the cuts from each subproblem are aggregated into a single cut. The advantages and disadvantages of both approaches are discussed in Chapter~5 of \citep{birge2011introduction}. It should be noted that whenever an optimization problem is a master problem, integrality is enforced on $x_{ijbt}^{\text{p},s}$. However, when moving up the scenario tree and the MP becomes a subproblem, the problem is relaxed to a linear program to enable the computation of dual variables. The subproblem is then,
\begin{subequations}\label{eq:spevmpc}
\begin{align}
        &\minimize_{x_{ijbt}^{\text{r},s}, x_{ijbt}^{\text{p},s}, x_{iibt}^{\text{c},s}, x_{iibt}^{\text{d},s}} 
        J_N(x_{ijbt}^{\text{r},s}, x_{ijbt}^{\text{p},s}, x_{iibt}^{\text{c},s}, x_{iibt}^{\text{d},s}) + \sum_{s=1}^S p_s \alpha_s\\
        & \text{subject to} \notag \\
        & \eqref{eq:flowcons},\ \eqref{eq:im},\ \eqref{eq:charger},\ \eqref{eq:non-antici1} \\
        & \alpha_s \geq Q^{s} + \sum_{ijb}\pi_{ijb}^{\text{r},s}(x_{ijb}^{\text{r},s}-\bar{x}_{ijb}^{\text{r},s}))+\pi_{ijb}^{\text{p},s}(x_{ijb}^{\text{p},s}-\bar{x}_{ijb}^{\text{p},s}))+\pi_{iib}^{\text{c},s}(x_{iib}^{\text{c},s}-\bar{x}_{iib}^{\text{c},s}))+\pi_{iib}^{\text{d},s}(x_{iib}^{\text{d},s}-\bar{x}_{iib}^{\text{d},s}))\label{eq:subBDcut} \\
        & \qquad \forall s \in S_{t+1} \\
        & x_{ji(b+\Delta b_{ji})(t-\Delta t_{ji})}^{\text{p}} = \bar{x}_{ji(b+\Delta b_{ji})(t-\Delta t_{ji})}^{\text{p}} \label{eq:barp}\\
        & x_{ji(b+\Delta b_{ji})(t-\Delta t_{ji})}^{\text{r}} = \bar{x}_{ji(b+\Delta b_{ji})(t-\Delta t_{ji})}^{\text{r}} \label{eq:barr}\\
        & x_{ii(b-\Delta b^c)(t-1)}^{\text{c}} = \bar{x}_{ii(b-\Delta b^c)(t-1)}^{\text{c}}\label{eq:barc}\\
        & x_{ii(b+\Delta d)(t-1)}^{\text{d}} = \bar{x}_{ii(b+\Delta d)(t-1)}^{\text{d}}\label{eq:barv2g}\\
        & s_{ij(t-1)}=\bar{s}_{ij(t-1)} \label{eq:bars}\\
        &x_{ijbt}^{\text{p}},x_{ijbt}^{\text{r}},x_{iibt}^{\text{c}},x_{iibt}^{\text{d}} \in \mathbb{R}^+ \qquad \forall i,j \in \mathcal{N}, \forall t \in \mathcal{T}, \forall b \in \mathcal{B}.
\end{align}
\end{subequations}
where constraints \cref{eq:barc,eq:barr,eq:barp,eq:barv2g,eq:bars} are the solution from the previous time step MP. The SP have the BD cuts from previous iteration of the NB, \cref{eq:subBDcut}.  
\subsection{Improving Nested Benders Decomposition}

To accelerate the convergence of the NBD algorithm, we employ several enhancement techniques. First, subproblems corresponding to the same stage can be solved in parallel, which can significantly reduce computational time and particularly when stage aggregation is used \citep{dempster1998parallelization}. Aggregating multiple stages reduces the depth of the decomposition tree and shifts computational effort from cut generation to problem solving. While solving each aggregated subproblem becomes more time-consuming, the overall number of Benders cuts required typically decreases, leading to faster convergence.

So-called warm-up cuts are implemented prior to starting the NBD algorithm. These cuts are based on simple heuristics. For each scenario, the vehicle imbalance at every station and time step is computed in the absence of rebalancing. Based on these imbalances, slack-penalizing constraints are added to the master problems to incentivize early rebalancing and charging actions. The cost of the slack variable corresponds to the lost revenue from not serving a customer. These warm-up cuts help guide the initial iterations toward more balanced solutions and can significantly reduce convergence time.




\section{Mobility network simulations and sensitivity analysis}
\label{sec:sim}
We evaluate our approach in two high-fidelity case studies, San Francisco and Chicago, constructed from large-scale taxi trip records. We first assess the accuracy of the proposed travel-demand predictions on these data. Next, we analyze the computational performance of the Nested Benders Decomposition (NBD) used to solve the underlying stochastic optimization problems. We then benchmark the proposed stochastic and robust algorithm in the AMoDeus simulator \citep{amodeus}. Finally, we perform a sensitivity analysis to quantify how vehicle energy efficiency, battery capacity, fleet size, and charger availability affect system performance.

\subsection{Simulation environment}
We evaluate our proposed algorithm using the high-fidelity transport simulator AMoDeus \citep{amodeus}, an extension of the Multi-Agent Transport Simulation framework (MATSim) \citep{matsim} specifically designed for AMoD systems. Two urban contexts are studied: San Francisco and Chicago, both based on real-world taxi data.  

The San Francisco case study utilizes taxi data from 2008, encompassing 464,045 customer trips collected from 500 taxis \cite{c7j010}. For simulation, we selected Thursday, May 29, 2008 with a total travel demand of 11,453 trips and an average trip distance of 3.31 km. The Chicago scenario is based on taxi data collected between 2013 and 2023, representing over 212 million trips \citep{chicago_taxi_data}. Our analysis focuses on Wednesday, July 24, 2019 which recorded a total travel demand of 32,406 trips and an average trip distance of 14.64 km, using data from July 15--24, 2019. To meet the respective demand levels, we simulate fleets of 500 vehicles in San Francisco and 1,600 in Chicago.

In both cities, vehicles are initially distributed across $\mathcal{N} = 10$ charging stations, each starting at 80\% SoC. The SoC is discretized into $\mathcal{B} = 50$ intervals. The simulation horizon is $T = 20$ time steps (200 minutes) with a time step length of $\Delta t = 10$ minutes, and operational planning is carried out over a robust horizon of 6 time steps (60 minutes). Vehicles are required to end the simulation with at least $b_{\text{end}} = 30\%$ SoC. An overview of all simulation parameters is provided in Table \ref{tab:simparameters}.

Vehicle battery capacities are set to $e^b = 40$ kWh for San Francisco and $e^b = 70$ kWh for Chicago. We use a non-linear charging curve derived from \citep{WASSILIADIS}, in which charging proceeds at a constant 100 kW until 45\% SoC and then decreases linearly to 30 kW at 95\% SoC. The stochastic optimization model applies scenario reduction per time step with levels [3, 3, 5, 5, 5], resulting in 1,125 total scenarios. Battery violation tolerance is set to $\epsilon^b = 0.2$ and demand violation tolerance to $\epsilon^t = 0.2$. Nested Benders Decomposition (NBD) is used with a gap tolerance of $10^{-2}$ and a maximum of 30 iterations. The terminal cost parameter $a_t$ is updated at each receding horizon iteration based on day-ahead electricity prices from Nordpool \citep{nordpool_prices}, using a lookahead window of $H = 36$ time steps (6 hours) and a baseline parameter $a_{\text{base}} = 0.10$~\euro/kWh.

A key enhancement to the simulation environment is the integration of an energy consumption estimation module, as AMoDeus does not include this functionality for transportation simulations. This module estimates energy consumption using detailed speed profiles based on the established model from \citep{basso2021electric}. Furthermore, to accurately represent real-world driving conditions, stochasticity was introduced into the simulation by modeling traffic light behavior and randomizing vehicle speeds. 

\begin{figure}
    \centering
    \includegraphics[width=\textwidth]{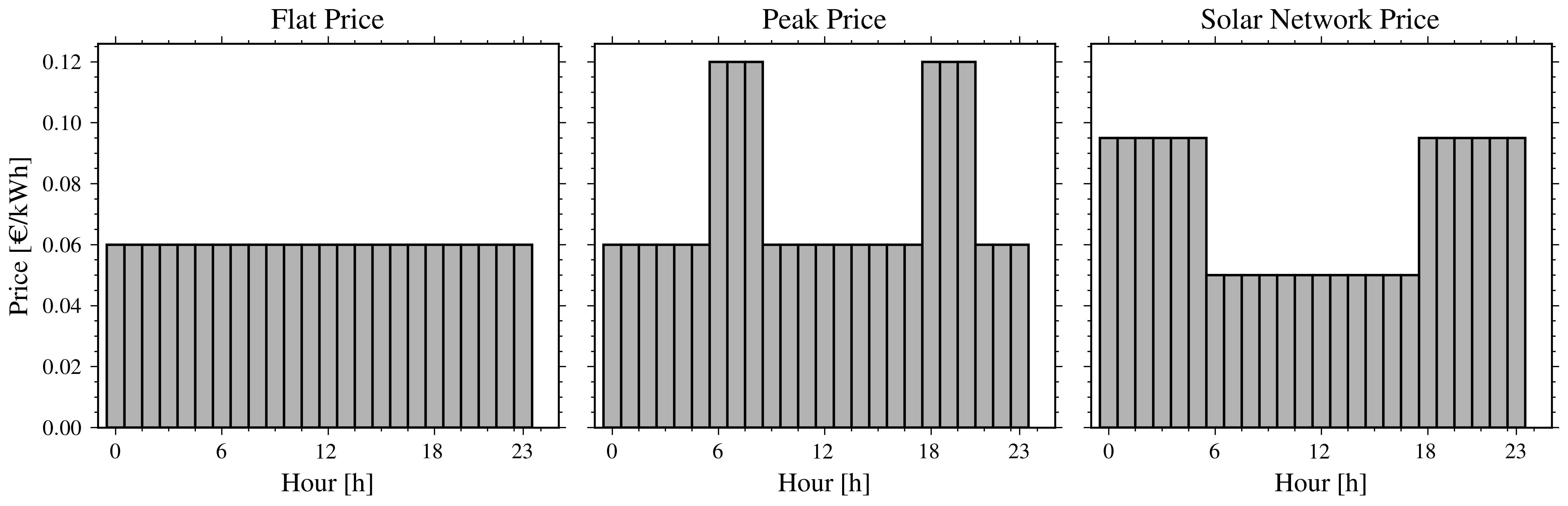}
    \caption{Electricity prices in €/kWh for three different pricing scenarios: Flat, Peak, and Solar Price, shown over a 24-hour period.}
    \label{fig:price_scenarios}
\end{figure}

For electricity pricing, we developed three distinct scenarios (Fig.\ref{fig:price_scenarios}): a flat-price profile, a peak-price profile, and a solar-price profile reflecting midday price drops from high solar generation, with the first two derived from Nordpool data \citep{nordpool_prices}. These represent daily operational variations rather than annual averages. For benchmarking the proposed algorithm, vehicle type vehicle 1 is used, while the sensitivity analysis compares three different vehicle types with varying energy consumption rates (Fig.\ref{fig:energyspeed}), under the peak-price scenario only.

All simulations were executed on a MacBook Pro with a 2.3 GHz Quad-Core Intel Core i7 processor and 16 GB of RAM. Optimization problems were solved using the Gurobi optimization library \citep{gurobi}.

\begin{table}
\centering
\caption{Simulation Parameters Overview}
\label{tab:simparameters}
\begin{tabular}{lll}
\hline
\multicolumn{3}{l}{\textbf{Common Parameters}} \\
\hline
Parameter & Value & \\
\hline
Number of stations ($\mathcal{N}$) & 10 & \\
Simulation horizon ($T$) & 20 time steps (200 minutes) & \\
Robust horizon & 6 time steps (60 minutes) & \\
Scenarios reduction per time step & [3,3,5,5,5] (1125 scenarios in total) & \\
Time step length ($\Delta t$) & 10 minutes & \\
Number of SoC intervals ($\mathcal{B}$) & 50 & \\
Minimum required SoC at end ($b_{\text{end}}$) & 30 \% & \\
V2G power ($P^d$) & 22 kW\\
Battery violation tolerance ($\epsilon^b$) & 0.2 & \\
Demand violation tolerance ($\epsilon^t$) & 0.2 & \\
Vehicle type & V1 \\
NBD gap tolerance & $10^{-2}$\\
Max number of BD iteration & 30 \\
\hline
\multicolumn{3}{l}{\textbf{City-Specific Parameters}} \\
\hline
Parameter & San Francisco & Chicago \\
\hline
Number of vehicles & 500 & 1600 \\
Battery capacity ($e^b$) & 40 kWh & 70 kWh \\
Total number of chargers & 100 & 200 \\
Total demand & 11,453 & 32,406 \\
Simulation date & May 29, 2008 & July 24, 2019 \\
\hline
\end{tabular}
\end{table}

\subsection{Travel Demand Prediction}

We conducted a comprehensive comparative analysis between BNF, our proposed prediction framework, and three established baselines: multi-output Gaussian Process regression with a spectral kernel (MOSK) \citep{parra2017spectral}, Long Short-Term Memory (LSTM) networks, and Vector AutoRegressive Moving Average (VARIMA) models \citep{zivot2006vector}. These baselines were selected based on their proven effectiveness in travel demand forecasting and represent a diverse set of statistical and machine learning approaches \citep{iglesias2018data, he2023data}. Like BNF, MOSK provides explicit uncertainty estimates, enabling a direct comparison of probabilistic forecasting performance. All models were evaluated on real-world datasets from San Francisco and Chicago. The BNF architecture used two hidden layers with 64 units each. Input features included time, origin coordinates (latitude and longitude), and destination coordinates. A Negative Binomial distribution was used as the prior predictive model to reflect the count nature of the data.

Model performance was assessed using two key metrics which were Root Mean Square Error (RMSE) and Mean Interval Score (MIS). RMSE quantifies the average deviation between predicted and actual values, with lower values indicating higher accuracy. MIS penalizes both the width and coverage of prediction intervals, providing a measure of uncertainty calibration; lower values correspond to sharper and more reliable probabilistic forecasts. Results are shown in Table~\ref{tab:rmse_mis_comparison}.

BNF achieved the best performance across all scenarios in both RMSE and MIS. In San Francisco, BNF reduced RMSE significantly by an average of 36.6\% compared to LSTM, 29.7\% compared to VARIMA, and 19.7\% compared to MOSK. The improvements were more pronounced in Chicago, with average RMSE reductions of 66.7\%, 47.1\%, and 15.8\% compared to LSTM, VARIMA, and MOSK, respectively. For uncertainty quantification, BNF also demonstrated a clear advantage, reducing MIS by an average of 42.7\% in San Francisco and 35.5\% in Chicago relative to MOSK.

In addition to demonstrating strong average performance, BNF showed greater robustness during peak demand periods (06:00–10:00 and 17:00–21:00), which are especially important for real-world EAMoD operations. During these high-demand windows, models such as MOSK exhibited significantly higher errors. For example, during the evening peak in Chicago, BNF reduced RMSE by 49.2\% compared to MOSK, indicating its superior ability to handle sharp fluctuations in travel demand. The substantial reductions in both RMSE and MIS contribute to more reliable scenario trees. A lower RMSE results in a more accurate average scenario, while a lower MIS improves the estimation of scenario probabilities within the scenario tree.

\begin{figure}
\centering
\includegraphics[width=\linewidth]{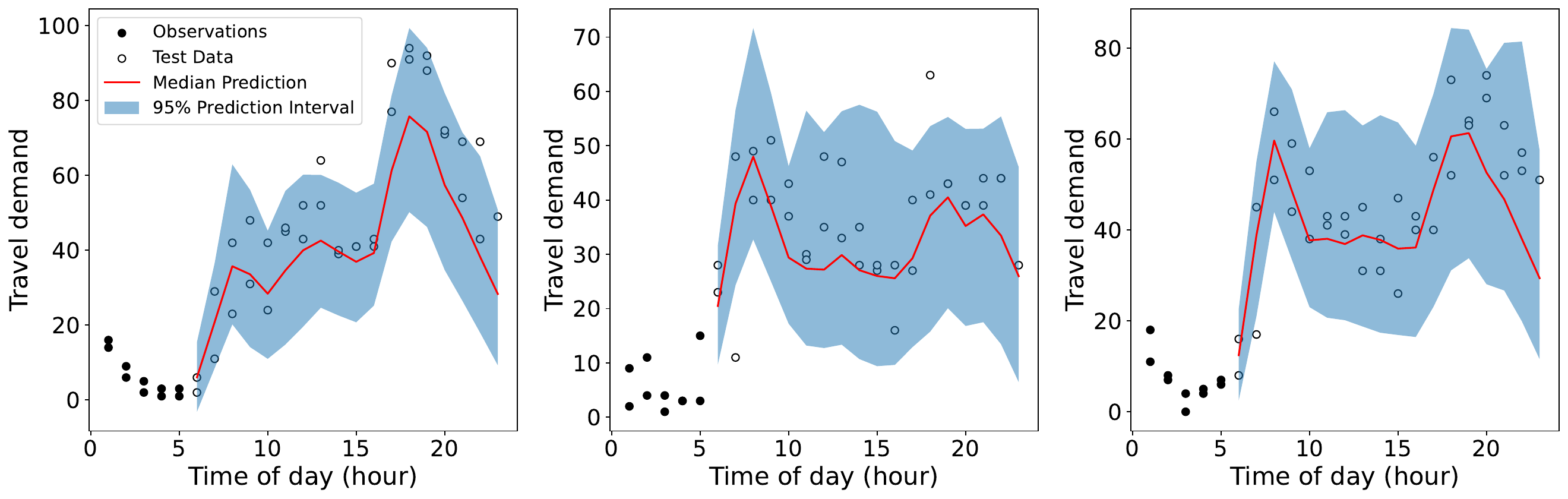}
\caption{Predicted travel demand with 95\% prediction intervals for selected locations. Black dots show training data, white dots indicate test data, the red line is the median prediction, and blue shaded areas represent the 95\% prediction intervals.}
\label{fig:travel_demand_predictions}
\end{figure}

As shown in Fig. \ref{fig:travel_demand_predictions}, BNF generates accurate travel demand forecasts with well-calibrated 95\% prediction intervals across selected Chicago pickup locations. While the median prediction does not perfectly match realized demand, the uncertainty estimates provide valuable additional information. This highlights the importance of considering predictive uncertainty in downstream decision-making and motivates the construction of scenario trees based on the probabilistic forecasts.

\begin{table}
    \centering
    \caption{Comparison of RMSE and MIS for different forecasting models in San Francisco and Chicago across key time periods.}
    \label{tab:rmse_mis_comparison}
    \begin{tabular}{llcccccc}
        \toprule
        Metric & Algorithm & \multicolumn{3}{c}{San Francisco} & \multicolumn{3}{c}{Chicago} \\
        \cmidrule(lr){3-5} \cmidrule(lr){6-8}
        & & 06:00--10:00 & 10:00--14:00 & 17:00--21:00 & 06:00--10:00 & 10:00--14:00 & 17:00--21:00 \\
        \midrule
        \multirow{4}{*}{RMSE}
        & LSTM   & 14.5175 & 18.8251 & 24.2956 & 27.4708 & 85.3836 & 133.4684 \\
        & VARIMA & 18.0840 & 9.6063  & 12.0309 & 29.3161 & 26.6069 & 35.5457 \\
        & MOSK   & 10.0874 & 7.8890  & 10.5079 & 26.0935 & 26.0016 & 28.8127 \\
        & BNF    & \textbf{8.1455} & \textbf{6.9751} & \textbf{8.0259} & \textbf{25.2362} & \textbf{13.6507} & \textbf{14.6357} \\
        \midrule
        \multirow{2}{*}{MIS}
        & MOSK   & 83.2670 & 63.8836 & 77.2634 & 193.0978 & 154.8194 & 148.6521 \\
        & BNF    & \textbf{40.0050} & \textbf{33.6349} & \textbf{53.0234} & \textbf{110.9026} & \textbf{98.2883} & \textbf{103.8081} \\
        \bottomrule
    \end{tabular}
\end{table}

\subsection{Nested Benders Decomposition Performance}
\label{sec:nbd_results}

We analyze the computational performance of Nested Benders Decomposition (NBD), which is used to solve the multistage stochastic optimization problems underlying our SMPC approach. We do not compare against the extensive form of the stochastic problem, as it becomes computationally intractable beyond very small instances. Instead, we evaluate NBD in terms of runtime, convergence behavior, and scalability with respect to the number of scenarios.  

In the nested Benders decomposition, one optimality gap is computed for each master problem and subproblem across the scenario tree. Since these gaps can vary considerably across nodes, we report both the maximum and the median gap as indicators of convergence quality. The maximum gap highlights the worst-case convergence behavior across all nodes, while the median gap reflects the typical convergence performance and is less sensitive to outliers. The optimality gap reported here refers specifically to the Benders decomposition procedures. In each iteration, the Benders cuts provide a lower bound on the subproblem cost for the current master solution, while the incumbent solution provides an upper bound. The relative gap is measured as
\[
\text{Gap} \;=\; \frac{UB - LB}{UB} \times 100\% ,
\]
where $UB$ denotes the incumbent upper bound and $LB$ the lower bound from the cuts. In the ideal case, this gap equals $0\%$, which means that the decomposition have reached the exact optimum. However, driving the gap fully to zero can be computationally prohibitive. In practice, we terminate the NBD algorithm if the gap does not improve for five consecutive iterations or if the runtime exceeds 600 seconds. The reported gaps reflect only the internal convergence of the decomposition. Since the subproblems are solved in relaxed form, the true duality gap with respect to the extensive formulation remains unobservable without solving the full problem.

Table~\ref{tab:runtime-scaling} reports the runtime, number of iterations, and the relative optimality gaps across three scenario set sizes. For the smallest instance (135 scenarios), NBD converges within 10 iterations and achieves a near-exact solution with a final max gap of 0.08\%. Increasing the number of scenarios to 540 raises the runtime to about 90 seconds, but convergence is still reached in fewer than 10 iterations, with a small median gap of 0.09\%.  

For the largest instance (1035 scenarios), the runtime increases to nearly 6 minutes and the number of iterations to 15. While the maximum gap observed during the iterations is higher (4.52\%), the algorithm ultimately converges to a high-quality solution with a median gap of 1.50\%. These results demonstrate that, although runtime grows with the number of scenarios, NBD remains computationally tractable and consistently delivers solutions with small optimality gaps, even for problem sizes where solving the extensive form is impractical.

\begin{table}
  \centering
  \caption{Runtime, number of iterations, and relative optimality gaps of NBD across different scenario counts.}
  \label{tab:runtime-scaling}
  \begin{tabular}{rrrrr}
    \toprule
    \textbf{Scenarios} & \textbf{Runtime (sec)} & \textbf{Iterations} & \textbf{Max Gap (\%)} & \textbf{Median Gap (\%)} \\
    \midrule
    135 & 17.15 & 10 & 0.08 & 0.08 \\
    540 & 90.26 & 9 & 0.79 & 0.09 \\
    1035 & 355.55 & 15 & 4.52 & 1.50 \\
    \bottomrule
  \end{tabular}
\end{table}

\subsubsection{Sensitivity to scenario tree configuration}
To evaluate the impact of scenario tree design on operational performance, we conducted a sensitivity analysis varying both the total number of scenarios and their allocation across stages.
We compared two allocation strategies with equal total scenarios: (i)~early branching [5,5,5,3,3], which allocates more scenarios to early stages to resolve near-term uncertainty; and (ii)~late [3,3,5,5,5], our baseline configuration, which concentrates branching in the later stages. Additionally, a medium configuration [3,3,3,4,5] with 540 total scenarios and a small configuration [3,3,3,5] with 135 total scenarios are included to assess the effect of decreasing the total scenario count. Table~\ref{tab:scenario_sensitivity} reports the results for San Francisco and Chicago.
The results show a modest but consistent improvement in tail performance as the total number of scenarios increases, while median wait times remain largely stable across configurations. In San Francisco, moving from the Small to the Large configuration leaves the median wait time unchanged at 229,s, while the 95th percentile decreases from 559,s to 548,s. In Chicago, the median similarly remains at 214,s, with the 95th percentile reducing from 783,s to 767,s, reflecting the higher sensitivity to uncertainty resolution under greater demand intensity. Beyond 1125 scenarios, gains plateau while computational cost increases substantially, justifying this as an appropriate upper bound.
For a fixed total of 1125 scenarios, the Late [3,3,5,5,5] configuration achieves a marginally lower 95th percentile wait time than the Early [5,5,5,3,3] configuration in both cities (548\,s vs.\ 550\,s in San Francisco; 767\,s vs.\ 767\,s in Chicago), with comparable total and rebalancing distances. This can be attributed to the heterogeneous nature of forecast uncertainty across the planning horizon: as shown in Fig.~\ref{fig:travel_demand_predictions}, prediction uncertainty is relatively low in the near-term stages and increases toward later stages. Consequently, allocating higher branching factors to early stages, where scenarios are nearly indistinguishable, yields little additional representational value, whereas denser branching in later stages better captures the growing spread in demand and charger availability. On this basis, we adopt the Late [3,3,5,5,5] configuration as the default, as concentrating branching in stages with greater uncertainty leads to a more informative scenario tree for a given computational budget.

\begin{table}
    \centering
    \renewcommand{\arraystretch}{1.2}
    \caption{Sensitivity analysis of scenario tree configuration for San Francisco (fleet size 500) and Chicago (fleet size 1600). Metrics include median and 95\% quantile wait times (WT), total distance and rebalancing distance per vehicle.}
    \begin{tabular}{llrrrrr}
        \toprule
        \textbf{City} & \textbf{Configuration} & \textbf{Scenarios} & \textbf{Median WT} & \textbf{95\% WT} & \textbf{Total Dist./Veh.} & \textbf{Reb. Dist./Veh.} \\
         & & (total) & [\si{s}] & [\si{s}] & [\si{km}] & [\si{km}] \\
        \midrule
        \multirow{5}{*}{Chicago}
        & Small $[3,3,3,5]$       & 135  & 214 & 783 & 495.2 & 69.1 \\
        & Medium $[3,3,3,4,5]$    & 540  & 214 & 767 & 500.7 & 70.1 \\
        & Large $[3,3,5,5,5]$     & 1125 & 214 & 767 & 500.7 & 70.1 \\
        & Early $[5,5,5,3,3]$     & 1125 &  214 & 767 & 506.3 & 69.8 \\
        \midrule
        \multirow{5}{*}{San Francisco}
        & Small $[3,3,3,5]$       & 135  & 229 & 559 & 412.4 & 103.7 \\
        & Medium $[3,3,3,4,5]$    & 540  & 229 & 555 & 413.2 & 106.3 \\
        & Large $[3,3,5,5,5]$     & 1125 & 229 & 548 & 414.1 & 108.5 \\
        & Early $[5,5,5,3,3]$     & 1125 & 229 & 550 & 412.5 & 107.5 \\
        \bottomrule
    \end{tabular}
    \label{tab:scenario_sensitivity}
\end{table}

\subsection{EAMoD Optimization Performance Comparison}

We benchmark our proposed SMPC algorithm, which explicitly accounts for uncertainty in predictions, against three alternative control strategies:
\begin{itemize}
    \item \textbf{Deterministic Mean Prediction Optimization (DMP):} Uses only the expected values of predicted travel demand, charging station occupancy, travel time, and energy consumption from BNF, without considering uncertainty.
    \item \textbf{Robust Chance-Constrained Optimization (RCC):} A robust approach inspired by \citep{jacobsen2023predictive}, designed to guarantee performance under worst-case prediction errors.
    \item \textbf{Reactive Baseline (RB):} A reactive strategy with no forecasting. Vehicles are charged whenever the state of charge (SoC) falls below 20\%, up to 80\%. No vehicle-to-grid (V2G) operations are considered.
\end{itemize}
All controllers are re-optimized every 10 minutes using a receding horizon approach. The results for Chicago and San Francisco are summarized in Tables ~\ref{tab:combined_wt} and ~\ref{tab:combined_price_v2g}.

In Chicago, SMPC achieves the lowest passenger delays. The median waiting time is 214~s, essentially the same as for DMP (215~s), but significantly lower at the 95th percentile with 767~s compared to 872~s for DMP (12\% improvement), 822~s for RCC, and 955~s for the reactive baseline (20\% improvement). Operationally, SMPC requires 70.1~km of rebalancing per vehicle, which is 27\% less than RCC (96.6~km), while keeping the total distance traveled (500.7~km) close to RCC (510.1~km). RB travels the least overall distance (456.1~km), but this comes at the cost of much longer wait times.

In San Francisco, SMPC again dominates in terms of service quality. The median waiting time is 229~s, compared to 240~s for DMP (5\% higher), 254~s for RCC, and 360~s for the reactive baseline, corresponding to a 36\% improvement over RB. At the 95th percentile, SMPC records 548~s compared to 584~s for DMP, 559~s for RCC, and 670~s for RB, improving extreme waits by almost 20\% relative to the reactive baseline. Operationally, SMPC requires 108.5~km of rebalancing per vehicle compared to 124.8~km for RCC, a 13\% reduction, while keeping total distance traveled (414.1~km) lower than RCC (431.6~km). Although DMP uses slightly less rebalancing (99.5~km), SMPC achieves consistently lower waiting times across the distribution, showing that the modest increase in repositioning yields substantially improved service reliability.

Energy costs and V2G revenues show similar trends in both cities. Under flat tariffs, all methods incur €0.06/kWh in costs and generate no revenue, as expected. However, under peak and solar tariffs, SMPC and DMP consistently achieve the lowest effective energy costs and the highest revenues. In Chicago, SMPC earns almost 975 € under solar pricing, which is approximately double the value in San Francisco of €509, reflecting the larger fleet and heavier charging activity. RCC captures somewhat less revenue due to its conservative scheduling, while RB never participates in V2G and pays the highest electricity prices.

Overall, these results demonstrate that the SMPC model strikes a balance between service quality, operational efficiency and energy costs. It delivers shorter wait times than deterministic or reactive baselines, avoids the conservatism of robust optimization and captures meaningful grid revenues by exploiting dynamic electricity prices.
While the improvements over deterministic and reactive baselines are consistent 
across all metrics, the absolute performance gaps remain moderate. This is partly 
explained by the relatively short 10-minute control interval used in the SMPC 
framework. Frequent re-optimization introduces strong closed-loop feedback, 
allowing the system to correct forecast errors rapidly and thereby reducing the 
marginal benefit of long-horizon anticipation.

This highlights an important structural property of high-frequency receding 
horizon control: as the update frequency increases, feedback dominates 
forecasting accuracy, and the incremental value of multi-stage stochastic 
optimization decreases. In contrast, for longer control intervals—where 
decisions must be committed further in advance and corrective adjustments are 
delayed—the relative advantage of anticipatory stochastic optimization is 
expected to increase.

\subsubsection{SMPC Component Analysis}
To isolate the contribution of each component in the proposed framework, we conducted a component analysis study of the SMPC with four configurations: (1)~the full SMPC baseline; (2)~SMPC without chance constraints, where energy consumption and travel time are set to their expected values; (3)~SMPC without the scenario tree, using a single expected-value path; and (4)~SMPC with a reduced robust horizon of 2 stages. Table~\ref{tab:ablation_sf_peak} reports operational performance for San Francisco under the peak pricing scenario.

The scenario tree is the most impactful component: removing it increases the median wait time from 229s to 240s and the 95th percentile from 548s to 571s, closely matching the DMP baseline, which confirms that branching over demand scenarios is the key differentiating element of the framework. Removing chance constraints leaves the median unchanged but degrades the 95th percentile (548→568s), indicating that chance constraints primarily improve tail reliability by preventing dispatch of vehicles with insufficient charge. Reducing the robust horizon to 2 stages produces a moderate but consistent degradation across both metrics (median 229→236s, 95\% 548→561s), demonstrating graceful degradation with a shorter planning horizon while retaining most of the performance gain. Together, these results confirm that each component contributes distinctly: the scenario tree drives average performance, chance constraints improve service reliability, and the robust horizon balances solution quality with computational tractability.

\begin{table}[h]
    \centering
    \renewcommand{\arraystretch}{1.2}
    \caption{SMPC Component Analysis (San Francisco, peak pricing).}
    
    \begin{tabular}{llrrrr}
        \toprule
        \textbf{Configuration} & \textbf{Method} & \textbf{Median} & 
        \textbf{95\%} & \textbf{Total Dist./Veh.} & 
        \textbf{Reb. Dist./Veh.} \\
         & & [\si{s}] & [\si{s}] & [\si{km}] & [\si{km}] \\
        \midrule
        \multirow{4}{*}{San Francisco}
        & Full SMPC                      & 229 & 548 & 414.1 & 108.5 \\
        & No chance constraints          & 229 & 568 & 410.6 & 103.3 \\
        & No scenario tree (single path) & 240 & 571 & 408.3 & 101.7 \\
        & Robust horizon 2               & 236 & 561 & 410.3 & 102.5 \\
        \bottomrule
    \end{tabular}
    \label{tab:ablation_sf_peak}
\end{table}


\begin{table}
    \centering
    \renewcommand{\arraystretch}{1.2}
        \caption{Operational performance in Chicago (fleet size 1600) and San Francisco (fleet size 500) under four control strategies. Metrics include mean, median (50\% quantile), and 95\% quantile wait times (WT), as well as total and rebalance distance per vehicle.}
    \begin{tabular}{llrrrr}
        \toprule
        \textbf{City} & \textbf{Method}  & \textbf{Median} & \textbf{95\%} & \textbf{Total Dist./Veh.} & \textbf{Reb. Dist./Veh.} \\
         &  & [\si{s}] & [\si{s}] & [\si{km}] & [\si{km}] \\
        \midrule
        \multirow{4}{*}{Chicago} 
        & \textbf{RB}  & 272 & 955 & 456.1 & 4.6 \\
        & \textbf{DMP}  & 215 & 872 & 490.0 & 65.1 \\
        & \textbf{RCC}  & 220 & 822 & 510.1 & 96.6 \\
        & \textbf{SMPC} & 214 & 767 & 500.7 & 70.1 \\
        \midrule
        \multirow{4}{*}{San Francisco} 
        & \textbf{RB}  & 360 & 670 & 401.1 & 3.8 \\
        & \textbf{DMP} & 240 & 584 & 407.1 & 99.5 \\
        & \textbf{RCC}  & 254 & 559 & 431.6 & 124.8 \\
        & \textbf{SMPC}  & 229 & 548 & 414.1 & 108.5 \\
        \bottomrule
    \end{tabular}
    \label{tab:combined_wt}
\end{table}

\begin{table}
    \centering
    \renewcommand{\arraystretch}{1.2}
        \caption{Average electricity price paid per kilowatt-hour (€/kWh) and corresponding total fleet vehicle-to-grid (V2G) revenue (€) under flat, peak, and solar pricing schemes for Chicago and San Francisco.}
    \begin{tabular}{llrrrrrr}
        \toprule
        \textbf{City} & \textbf{Method}  
        & \textbf{Flat Price} & \textbf{Flat V2G Rev.} 
        & \textbf{Peak Price} & \textbf{Peak V2G Rev.} 
        & \textbf{Solar Price} & \textbf{Solar V2G Rev.} \\
        &  & [€/kWh] & [€] & [€/kWh] & [€] & [€/kWh] & [€] \\
        \midrule
        \multirow{4}{*}{Chicago} 
        & \textbf{RB}   & 0.06 & 0.0 & 0.083 & 0.00 & 0.09 & 0.00 \\
        & \textbf{DMP}  & 0.06 & 0.0 & 0.078 & 565 & 0.07 & 981 \\
        & \textbf{RCC}  & 0.06 & 0.0 & 0.076 & 547 & 0.08 & 944 \\
        & \textbf{SMPC} & 0.06 & 0.0 & 0.075 & 563  & 0.07 & 975 \\
        \midrule
        \multirow{4}{*}{San Francisco} 
        & \textbf{RB}   & 0.06 & 0.0 & 0.08 & 0.00 & 0.090 & 0.00 \\
        & \textbf{DMP}  & 0.06 & 0.0 & 0.07 & 290 & 0.070 & 515 \\
        & \textbf{RCC}  & 0.06 & 0.0 & 0.08 & 285 & 0.074 & 453 \\
        & \textbf{SMPC} & 0.06 & 0.0 & 0.069 & 291 & 0.068 & 509 \\
        \bottomrule
    \end{tabular}
    \label{tab:combined_price_v2g}
\end{table}
\subsection{Experimental setup for sensitivity analysis}
\begin{figure}
    \centering
    \includegraphics[width=0.5\linewidth]{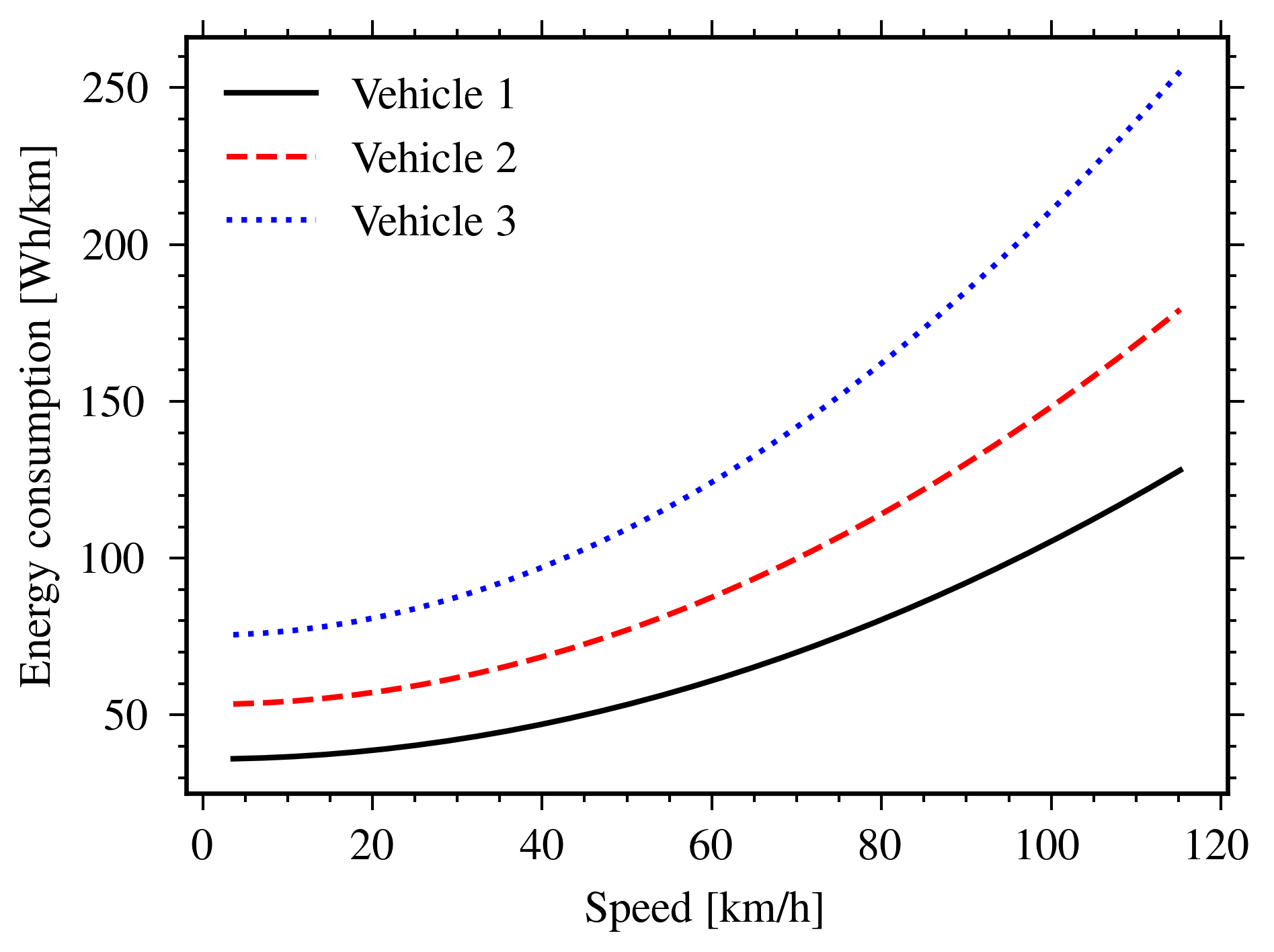}
    \caption{Energy consumption of the different vehicles without auxiliary power and with a 40 kWh battery.}
    \label{fig:energyspeed}
\end{figure}

This section investigates the impact of battery size and vehicle energy efficiency on the performance of urban EAMoD systems in San Francisco and Chicago. We analyze three distinct vehicle types, representing high, medium, and low energy efficiency, and consider three battery capacities: 40, 70, and 100~kWh (Table~\ref{tab:experimental_setup}). Vehicle parameters, including those determining efficiency, are summarized in Table~\ref{tab:vehatri}, based on the energy consumption model from \citep{basso2021electric}. Vehicle weights listed in the table exclude battery mass, which is added separately assuming a specific energy of 268~Wh/kg \citep{gunter2022state}. Fig.\ref{fig:energyspeed} illustrates the energy consumption rates of the three vehicle types with a 40~kWh battery, excluding auxiliary loads.

\begin{table}
\centering
\caption{Overview of the experimental setup for the mobility-on-demand simulations in San Francisco and Chicago. The different vehicle types corresponds to different energy efficiency (low, medium and high).}
\label{tab:experimental_setup}
\renewcommand{\arraystretch}{1.3}
\begin{tabular}{lll}
\hline
\textbf{Factor} & \textbf{San Francisco} & \textbf{Chicago} \\
\hline
Vehicle Type & 1 (low), 2 (medium), 3 (high) & 1 (low), 2 (medium), 3 (high) \\
Battery Size (kWh) & 40, 70, 100 & 40, 70, 100 \\
Fleet Size (vehicles) & 450, 500, 550 & 1400, 1600, 1800 \\
Number of Chargers & 100, 150, 200 & 200, 300, 400 \\
\hline
\end{tabular}
\end{table}

\begin{table}
\centering
\caption{Vehicle parameters used to model energy consumption. Symbols, descriptions, and units are listed in the first three columns.}
\label{tab:vehatri}
\begin{tabular}{l l l r r r}
\toprule
Symbol & Description & Unit & Vehicle 1 & Vehicle 2 & Vehicle 3 \\
\midrule
$m$ & Weight & kg & 1500 & 1700 & 1900 \\
$\eta^+$ & Efficiency (acceleration) & – & 0.90 & 0.85 & 0.80 \\
$\eta^-$ & Efficiency (deceleration) & – & 0.90 & 0.85 & 0.80 \\
$\eta^{\text{nomial}}$ & Efficiency (constant speed) & – & 0.90 & 0.85 & 0.80 \\
$c_r$ & Rolling resistance coefficient & – & 0.0072 & 0.009 & 0.0108 \\
$c_d$ & Drag coefficient & – & 0.22 & 0.25 & 0.30 \\
$p_{\text{aux}}$ & Auxiliary power & W & 800 & 900 & 1000 \\
$p_{\text{auxidle}}$ & Auxiliary idle power & W & 100 & 100 & 100 \\
$A$ & Frontal area & m$^2$ & 2.2 & 2.5 & 2.8 \\
\bottomrule
\end{tabular}
\end{table}

\subsection{Sensitivity analysis of battery size and energy efficiency (San Francisco)}
Battery capacity significantly influences system performance, though the extent depends on vehicle efficiency (Table \ref{tab:vehwtsf}). For Vehicle 1, the most energy-efficient option, wait times remain nearly constant across all battery sizes. Increasing the battery from 40 kWh (median 220s) to 100 kWh (median 219s) yields virtually no change in performance. Similarly, for vehicle 1 the 75th and 95th percentile wait times remain below 450s across all capacities. This indicates that Vehicle~1’s service quality is not constrained by battery size, even with the smallest pack.

Vehicles 2 and 3 are more sensitive to battery size. For Vehicle 2, the median wait time decreases slightly from 225 s at 40 kWh to 217 s at 70 kWh, while the share of passengers waiting more than 600 s declines from 154 to 151. Increasing the battery to 100 kWh provides only marginal improvements, with the median wait time rising slightly to 219 s and the number of passengers waiting more than 600 s falling to 144. This suggests that a 70 kWh battery is sufficient to mitigate charging as a performance bottleneck for Vehicle 2, with further increases offering diminishing returns.
For Vehicle 3, the least efficient option, the effect of battery size is far more pronounced. With 40 kWh, the system exhibits very poor performance, with a median wait time of 325s, a 95th percentile of 2625s, and more than 3600 passengers experiencing delays over 600s. Upgrading to 70 kWh brings dramatic improvements: the median drops to 219s, the 95th percentile to 432s, and delays over 600s fall by more than 95\% to just 145. Increasing to 100 kWh does not significantly improve service further, with a median of 222s and 149 delays over 600s. This highlights that for inefficient vehicles, battery size is a critical driver of feasibility, with performance plateauing around 70~kWh.

These results yield two key insights. First, energy-efficient vehicles like Vehicle~1 exhibit lower sensitivity to battery size, enabling effective performance with smaller, lighter battery packs. This directly contributes to overall system efficiency and potentially lower operational costs. Second, while less efficient vehicles require larger batteries to achieve acceptable service quality, the performance gains tend to plateau around 70~kWh. Importantly, these service dynamics also shape revenues from V2G services. Efficient vehicles consistently generate higher V2G revenues, since their lower consumption leaves more capacity for grid interaction. For example, Vehicle 1 earns €304 per vehicle at 40 kWh and rises to €390 at 100 kWh, the highest across all types. Vehicle 3, by contrast, produces the lowest revenues (e.g., €260 at 40 kWh, €360 at 100 kWh), as its higher driving demand limits grid availability. Larger batteries generally boost revenues across all vehicles, but the marginal gains diminish at the upper end: Vehicle 2 increases from €285 at 40 kWh to €342 at 70kWh, while the additional jump to €386 at 100 kWh is relatively small. Thus, the same diminishing returns seen in wait times and charging behavior are mirrored in V2G participation. 

\begin{figure}
    \centering
    \begin{subfigure}[b]{0.48\textwidth}
        \centering
        \includegraphics[width=\textwidth]{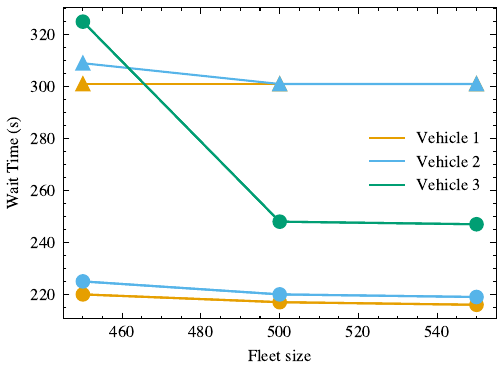}
        \caption{Median (circle) and 75th percentile (triangle) of wait times across fleet size}
        \label{fig:waittime_median_fleet}
    \end{subfigure}
    \hfill
    \begin{subfigure}[b]{0.48\textwidth}
        \centering
        \includegraphics[width=\textwidth]{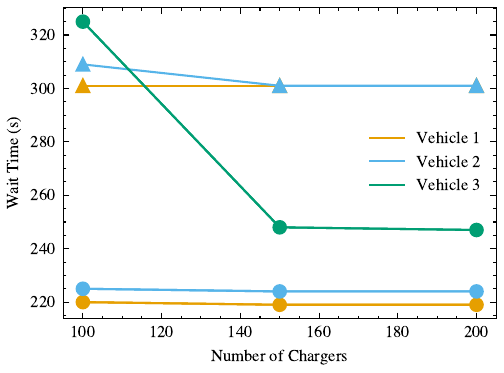}
        \caption{Median (circle) and 75th percentile (triangle) of wait times across number of chargers}
        \label{fig:waittime_median_charger}
    \end{subfigure}
    \caption{Comparison of wait time metrics in San Francisco.  Fig. a shows metrics across different fleet sizes and vehicle types with 100 chargers. Fig. b shows metrics across different number of charger and vehicle types with a fleet size of 450 vehicles. All plots assume a 40 kWh battery capacity. For clarity, 75th percentile wait times for Vehicle 3 are not shown in the plots, but their values are as follows: for varying \textit{fleet size} (450, 500, 550 vehicles), the Q75 values are \textbf{1129\,s}, \textbf{610\,s}, and \textbf{590\,s}, respectively; for varying \textit{number of chargers} (100, 150, 200), the Q75 values are \textbf{1129\,s}, \textbf{654\,s}, and \textbf{650\,s}, respectively.}
    \label{fig:waittime_combined}
\end{figure}

\begin{figure}
    \centering
    \begin{subfigure}[b]{0.48\linewidth} 
        \centering
        \includegraphics[width=\linewidth]{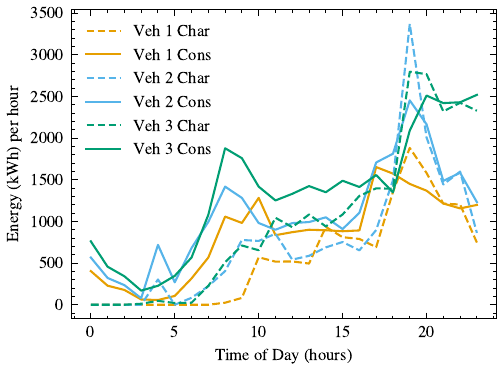}
        \caption{}
        \label{fig:energy-charged_sf} 
    \end{subfigure}
    \hfill 
    \begin{subfigure}[b]{0.48\linewidth} 
        \centering
        \includegraphics[width=\linewidth]{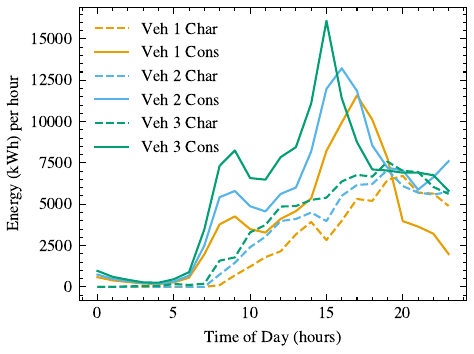}
        \caption{}
        \label{fig:energy-charged_chicago} 
    \end{subfigure}
    \caption{Comparison of hourly energy charged and consumed across three different vehicle types with 40 kWh battery capacity for (a) San Francisco and 70 kWh for (b) Chicago. Each line represents the total energy (in kWh) aggregated across all vehicles in a fleet. Charging events are shown with dashed lines, and consumption with solid lines.}
    \label{fig:energy-charged}
\end{figure}

These findings are consistently reflected in other system-level metrics. Figures~\ref{fig:waittime_combined}  demonstrate that increasing fleet size and charger availability effectively reduce wait times, particularly for the less efficient Vehicle 3 and some for Vehicle 2. Vehicle 1, however, continues to exhibit stable performance, characterized by lower 75th percentile and minimal sensitivity to infrastructure constraints. Furthermore, Vehicle~3 has the highest energy consumption and consequently the greatest charging demand, especially during peak hours, see Fig. \ref{fig:energy-charged_sf}. Conversely, Vehicle~1 maintains a smoother, more aligned charging-consumption profile throughout the day.

\begin{figure}
    \centering
    \begin{subfigure}[b]{0.48\linewidth} 
        \centering
        \includegraphics[width=\linewidth]{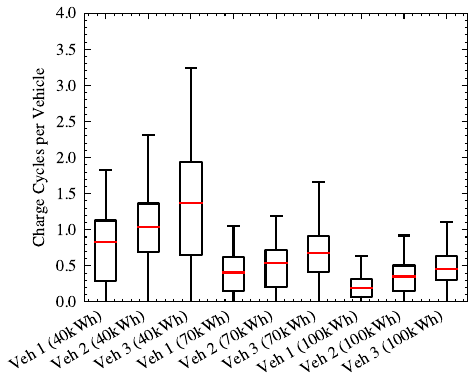}
        \caption{}
        \label{fig:chargecycles_sf} 
    \end{subfigure}
    \hfill 
    \begin{subfigure}[b]{0.48\linewidth} 
        \centering
        \includegraphics[width=\linewidth]{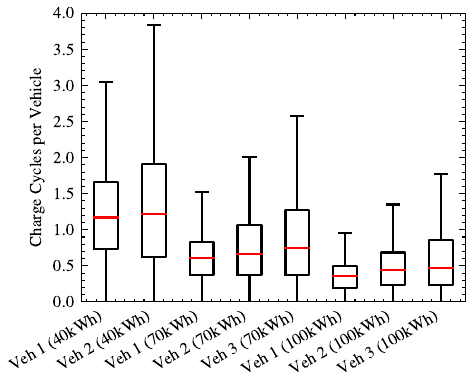}
        \caption{}
        \label{fig:chargecycles_chicago} 
    \end{subfigure}
    \caption{Observed charge cycles (20--80\% Depth of Discharge) for different vehicle types (1, 2, 3) and battery sizes (40, 70, 100 kWh) in (a) San Francisco and (b) Chicago. These distributions highlight the combined influence of vehicle characteristics and geographical location on battery charging patterns.}
    \label{fig:combined_chargecycles}
\end{figure}

Fig.\ref{fig:chargecycles_sf} reports the estimated number of daily charge cycles (defined as charging from 20\% to 80\% state of charge) across vehicle types and battery sizes. As anticipated, larger batteries consistently require fewer charging cycles. Vehicle 1 consistently exhibits the lowest and most stable charging frequency, with a median of just under one cycle per day even with a 40 kWh battery. In contrast, Vehicles 2 and 3 demonstrate higher and more variable charging frequencies, particularly with smaller battery configurations. For instance, Vehicle 3 with a 40~kWh pack exhibits a median of just under 1.5 cycles per day and substantial variability in its charging behavior. Increasing the battery capacity to 70~kWh or 100~kWh nearly halves the median number of cycles and significantly narrows the distribution of charging events.

These observed differences in charging frequency directly correlate with wait time performance. Frequent charging, especially for Vehicle~3 with small batteries, leads to service delays as vehicles become temporarily unavailable for passenger service. Larger batteries effectively mitigate both the number of charge cycles and the occurrence of such extended delays.

In summary, this analysis highlights that efficient vehicles can deliver consistent service with smaller batteries, while less efficient vehicles necessitate larger battery packs to avoid degraded performance. Reducing charge cycle frequency is paramount not only for prolonging battery lifespan but also for ensuring stable and responsive EAMoD service. These results underscore that optimal vehicle selection for urban EAMoD systems must consider not only immediate performance metrics but also long-term operational sustainability. Ultimately, if a vehicle is sufficiently energy efficient, a 40~kWh battery offers a compelling trade-off between cost, performance, and battery longevity.

\begin{table}
\centering
\caption{Summary of wait time and per-vehicle distance by battery capacity and vehicle type for San Francisco. The ``Wait $>$ 600s'' column reports the number of trips with wait times exceeding 600~s and the corresponding percentage of total trip requests ($n = 11{,}453$).}
\label{tab:vehwtsf}
\begin{tabular}{lrrrrrrrr}
\toprule
Vehicle &  SoC &       Median &          75\% &          95\% &  Wait $>$ 600s [count (\%)] &  TotDist/Veh & RebalDist/Veh & V2G Total \\
 & [kWh] & [s] & [s] & [s] &  & [km] & [km] & [\euro] \\
\midrule
      1 &   40 &          220 &          301 &          440 &         \textbf{137 (1.2\%)} &          240 &            79 & 304\\
      2 &   40 &          225 &          309 &          452 &         154 (1.3\%) &          206 &            42 & 285 \\
      3 &   40 &          325 &         1129 &         2625 &        3643 (31.8\%) &          216 &            37 & 260\\
      1 &   70 &          219 &          \textbf{296} &          432 &         145 (1.3\%) &          284 &           123 & 347 \\
      2 &   70 &   \textbf{217} &          \textbf{296} &          434 &         151 (1.3\%) &          241 &            80 & 342 \\
      3 &   70 &          219 &          299 &          \textbf{432} &         145 (1.3\%) &          209 &            47 & 316\\
      1 &  100 &          219 &          303 &          443 &         134 (1.2\%) &          291 &           129 & 390\\
      2 &  100 &          219 &          303 &          447 &         144 (1.3\%) &          292 &           130 & 386\\
      3 &  100 &          222 &          306 &          451 &         149 (1.3\%) &   \textbf{190} &   \textbf{32} & 360\\
\bottomrule
\end{tabular}
\end{table}

\subsection{Sensitivity analysis of battery size and energy efficiency (Chicago)}

We now examine how battery size and vehicle energy efficiency affect system 
performance in Chicago, which presents a different urban context from San 
Francisco. The same three vehicle types and battery capacities (40, 70, and 
100~kWh) are considered, except for Vehicle~3 with a 40~kWh battery, which 
is excluded because its limited capacity cannot sustain viable operation in 
Chicago.

Energy demand in Chicago is more concentrated during peak periods than in 
San Francisco, as shown in Fig.~\ref{fig:energy-charged_chicago}. Less 
efficient vehicles (Vehicle~3 in particular) exhibit a pronounced mismatch 
between energy consumption and charging, leading to steep late-day charging 
requirements. Vehicle~1 maintains the most balanced profile, with low and 
stable energy use throughout the day. This stability also allows Vehicle~1 
to contribute more consistently to V2G services, especially at larger 
battery sizes.

Battery size has a substantial influence on system performance in Chicago, 
with less efficient vehicles showing the greatest sensitivity to energy 
storage capacity. As reported in Table~\ref{tab:vehwtchicago}, Vehicle~1 
delivers consistently low wait times across all battery sizes, even under 
Chicago's higher demand. The median wait time improves modestly from 214~s 
at 40~kWh to 203~s at both 70 and 100~kWh. The effect is more visible in 
the upper percentiles: the 75th percentile drops from 329~s to 258~s 
between 40 and 70~kWh and remains unchanged at 100~kWh, indicating that a 
moderate battery increase is sufficient to reduce occasional peak delays. 
The 95th percentile falls from 1302~s at 40~kWh to 673~s at both 70 and 
100~kWh, cutting the longest delays by nearly half. The number of trips 
with wait times exceeding 600~s is likewise more than halved, dropping from 
4392 to approximately 1900. These results show that while Vehicle~1 already 
performs well at the median, larger batteries primarily improve service 
reliability by reducing infrequent but severe delays. Vehicle~1 also 
achieves the highest V2G revenues, ranging from \euro{}572 at 40~kWh to 
\euro{}580 at 100~kWh, confirming that efficiency and moderate battery 
sizing jointly support both passenger service and grid interaction.

Vehicle~2, by contrast, depends much more strongly on battery size. Its 
median wait time decreases from 293~s at 40~kWh to 221~s at 70~kWh and 
204~s at 100~kWh. The tail behaviour is even more striking: the 75th 
percentile shrinks from 2129~s at 40~kWh to 426~s at 70~kWh and 266~s at 
100~kWh, while the 95th percentile drops from 9772~s to 2919~s and then to 
692~s. This represents a shift from highly unstable to near-stable 
performance. The number of trips exceeding 600~s falls from 12\,176 at 
40~kWh to 6606 at 70~kWh and 2004 at 100~kWh. Although more efficient 
than Vehicle~3, Vehicle~2 remains sensitive to energy availability and 
benefits considerably from battery upgrades. Its V2G revenues, however, 
stay nearly flat across capacities (\euro{}563 at 40--70~kWh, \euro{}562 at 
100~kWh), indicating that most of its stored energy is consumed by travel 
demand rather than made available for grid services.

Vehicle~3, the least efficient type, is the most sensitive to battery 
capacity and consistently struggles under Chicago's high demand. At 
70~kWh, its median wait time is 316~s, but the upper tail reveals severe 
congestion: the 75th percentile reaches 5652~s and the 95th percentile 
exceeds 18\,000~s, with over 13\,000 trips delayed by more than 10 
minutes. A 100~kWh battery offers partial relief, reducing the median to 
237~s, the 75th percentile to 1072~s, and the 95th percentile to 
10\,830~s. Long-wait instances drop to 9912, but this remains far too 
high for acceptable service. These figures illustrate the fundamental 
limitations of low-efficiency vehicles in dense urban settings: even 
substantial battery investments cannot close the performance gap with more 
efficient alternatives. Vehicle~3 also generates the lowest V2G revenues 
(\euro{}559 at 70~kWh and \euro{}550 at 100~kWh), as its high consumption 
leaves little surplus energy for grid participation.

\begin{figure}
    \centering
    \begin{minipage}[b]{0.48\textwidth}
        \centering
        \includegraphics[width=\textwidth]{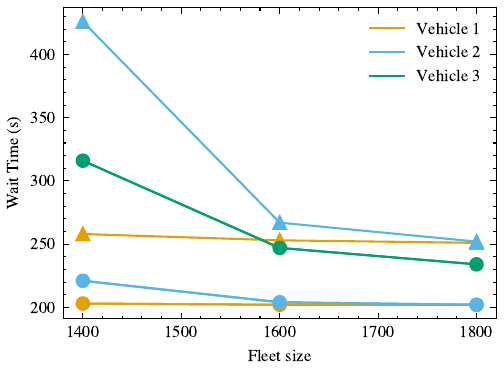}
        \textbf{(a)} Median (square) and 75th percentile (triangle) of wait times across number of chargers 
    \end{minipage}
    \hfill
    \begin{minipage}[b]{0.48\textwidth}
        \centering
        \includegraphics[width=\textwidth]{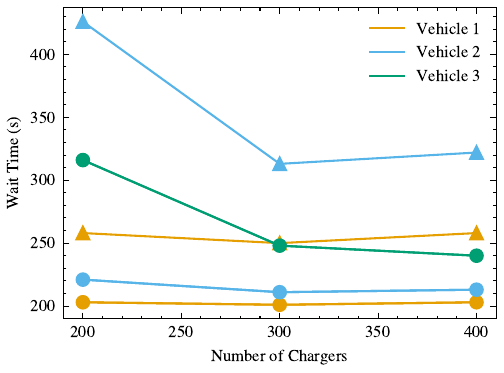}
        \textbf{(b)} Median (square) and 75th percentile (triangle) of wait times across number of chargers
    \end{minipage}
\caption{
Comparison of wait time metrics in Chicago. Fig. a shows metrics across different fleet sizes and vehicle types with 200 chargers. Fig. b shows metrics across different number of charger and vehicle types with a fleet size of 1400 vehicles. All plots assume a 70\,kWh battery capacity. For clarity, 75th percentile wait times for Vehicle 3 are not shown in the plots, but their values are as follows: for varying \textit{fleet size} (1400, 1600, 1800 vehicles), the Q75 values are \textbf{5651\,s}, \textbf{2180\,s}, and \textbf{969\,s}, respectively; for varying \textit{number of chargers} (200, 300, 400), the Q75 values are \textbf{5651\,s}, \textbf{1172\,s}, and \textbf{932\,s}, respectively.
}
\label{fig:waittime_combined_chicago}
\end{figure}

Fig.~\ref{fig:waittime_combined_chicago} shows how infrastructure scale 
affects performance. As fleet size increases from 1400 to 1800, all 
vehicles benefit, but the reduction is most pronounced for Vehicles~2 
and~3. Vehicle~3, for instance, sees its median wait time fall from over 
300~s to around 240~s. Similarly, increasing the number of chargers from 
200 to 400 lowers both median wait times and the 75th percentile, with 
the largest gains again accruing to less efficient vehicles.

Charging frequency data (Fig.~\ref{fig:chargecycles_chicago}) follow 
similar trends as in San Francisco. Vehicle~1 requires just over one 
cycle per day even with a 40~kWh battery, whereas Vehicles~2 and~3 charge 
more frequently and with greater variability, especially at smaller battery 
sizes. Vehicle~2 at 40~kWh has a median of about 1.2 cycles per day, 
dropping to around 0.7 at 70~kWh. Vehicle~3 consistently records the 
highest cycle counts, reflecting its greater energy demand. Frequent 
charging not only limits vehicle availability for passenger trips but also 
reduces the window for stable V2G participation.

The Chicago results reinforce the San Francisco findings while exposing 
sharper contrasts due to higher demand intensity. Vehicle~1 remains robust 
across battery sizes and system configurations, whereas Vehicles~2 and~3 
require both larger batteries and more infrastructure to maintain 
acceptable service levels. Pairing efficient vehicles with moderate battery 
capacities yields strong, scalable performance; less efficient platforms 
demand combined investment in capacity and infrastructure to avoid service 
degradation. From a V2G perspective, Vehicle~1 is again the most 
attractive option, combining stable service with the highest revenues. 
Vehicles~2 and~3 offer limited V2G potential because their stored energy 
is largely consumed by travel demand, confirming that vehicle efficiency 
is the primary determinant of grid-service opportunities in high-demand 
urban environments.

\begin{table}
\centering
\caption{Summary of wait time and per-vehicle distance by battery capacity and vehicle type for Chicago. The ``Wait $>$ 600s'' column reports the number of trips with wait times exceeding 600~s and the corresponding percentage of total trip requests ($n = 32{,}406$).}
\label{tab:vehwtchicago}
\begin{tabular}{lrrrrrrrrrr}
\toprule
Vehicle & SoC & Median & 75\% & 95\% & Wait $>$ 600s [count (\%)] & TotDist/Veh & RebalDist/Veh & V2G Total \\
 & [kWh] & [s] & [s] & [s] &  & [km] & [km] & [\euro] \\
\midrule
1 & 40 & 214 & 329 & 1302 & 4392 (13.6\%) & 512.3 & 67.1 & 572  \\
2 & 40 & 293 & 2129 & 9772 & 12176 (37.6\%) & 461.7 & 52.1 & 563  \\
1 & 70 & \textbf{203} & \textbf{258} & \textbf{673} & \textbf{1911 (5.9\%)} & 500.7 & 70.1 & 563  \\
2 & 70 & 221 & 426 & 2919 & 6606 (20.4\%) & 510.0 & 60.6 & 563  \\
3 & 70 & 316 & 5652 & 18175 & 13113 (40.5\%) & \textbf{432.6} & \textbf{50.9} & 559  \\
1 & 100 & \textbf{203} & \textbf{258} & \textbf{673} & 1941 (6.0\%) & 492.0 & 63.5 & 580  \\
2 & 100 & 204 & 266 & 692 & 2004 (6.2\%) & 504.2 & 68.8 & 562  \\
3 & 100 & 237 & 1072 & 10830 & 9912 (30.6\%) & 466.4 & 52.8 & 550  \\
\bottomrule
\end{tabular}
\end{table}

\newpage
\section{Conclusion}  
This paper presented a combined stochastic and robust model predictive  control framework for electric autonomous mobility-on-demand (EAMoD)  systems under multiple sources of uncertainty. By integrating  Bayesian Neural Field based probabilistic forecasting with  multi-stage stochastic optimization and robust chance constraints,  the proposed framework jointly optimizes dispatch, rebalancing,  charging, and V2G decisions in a unified formulation.   

A key methodological contribution is the tailored Nested Benders  Decomposition algorithm, which exploits the scenario tree structure  to solve large-scale multi-stage mixed-integer problems with  temporally correlated uncertainties. This enables real-time  applicability despite the exponential growth of the extensive-form  formulation. Unlike standard SAA formulations that typically assume  stagewise independence, the proposed approach preserves temporal  dependence across demand and infrastructure uncertainties, which is  essential in EAMoD systems.  

Simulation results on San Francisco and Chicago demonstrate  consistent improvements over deterministic, reactive, and robust  baselines, including up to 36\% reduced passenger waiting times,  27\% lower rebalancing distance, and 35\% electricity cost savings  through V2G participation. The Bayesian Neural Field forecasting  model was shown to outperform established baselines in both point  accuracy and calibrated uncertainty quantification, providing the  probabilistic inputs essential for effective scenario tree  construction. Importantly, the analysis reveals a structural  insight: the relative value of multi-stage stochastic anticipation  depends on the control update frequency. When receding horizon  updates are frequent, closed-loop feedback mitigates forecast  errors, reducing but not eliminating the marginal benefit of  anticipatory optimization. For longer decision intervals or  stronger temporal coupling in uncertainties, the advantage of  stochastic control is expected to grow.  

Overall, the proposed framework provides a scalable and  computationally tractable pathway for integrating predictive  uncertainty modeling and energy-aware fleet control in  next-generation autonomous mobility systems. Future research may  explore scenario tree construction based on the nested  distance~\cite{pflug2010version}, which provides formal  approximation guarantees by accounting for the filtration structure  of the stochastic process. Additionally, future work will explore alternative approaches to the terminal cost formulation, including multi-scale time discretizations, alternative scenario sampling methods, and layered synchronization MPC strategies that couple short-horizon real-time control with longer-horizon planning, to further improve long-horizon foresight within the receding horizon framework. Finally, the framework could be  extended to endogenous pricing, network congestion effects, or  co-optimization of charging infrastructure and fleet composition.
\section{Acknowledgement}
This work was co-funded by Vinnova, Sweden through the project: Simulation, analysis and modeling of future efficient traffic systems. This work was in part supported by the Transport Area of Advance within Chalmers University of Technology. The computations were enabled by resources provided by the National Academic Infrastructure for Supercomputing in Sweden (NAISS) at Chalmers e-Commons partially funded
by the Swedish Research Council through grant agreement no 2022-06725.
\section{Declaration of generative AI and AI-assisted technologies in the writing process}
During the preparation of this work, the authors used ChatGPT and DeepL exclusively to improve the readability and language of the manuscript. The scientific content, ideas, and conclusions are entirely the authors’ own. After using these tools, the authors carefully reviewed and edited the text, and they take full responsibility for all content of the published article.

\printcredits
\newpage
\bibliography{cas-refs}


\bio{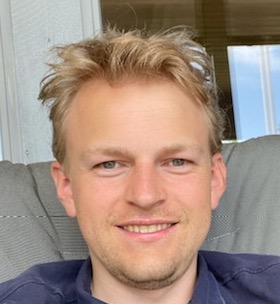}
\noindent \textbf{Sten Elling Tingstad Jacobsen} received the M.S. degree in applied physics from Chalmers University of Technology, Sweden, in 2020. He is currently working on his Ph.D. jointly at Volvo Cars AB and the Department of Electrical Engineering, Chalmers University of Technology, Sweden. His research interests include optimization, machine learning, control, and their application to transportation systems and the automotive area.
\endbio
\bio{figs/author3}
\textbf{Balázs Kulcsár} received the M.Sc. degree in traffic engineering and the Ph.D. degree from the Budapest University of Technology and Economics (BUTE), Budapest, Hungary, in 1999 and 2006, respectively. He has been a Researcher/Post-Doctor with the Department of Control for Transportation and Vehicle Systems, BUTE, the Department of Aerospace Engineering and Mechanics, University of Minnesota, Minneapolis, MN, USA, and the Delft Center for Systems and Control, Delft University of Technology, Delft, the Netherlands. He is currently a Professor with the Department of Electrical Engineering, Chalmers University of Technology, Gothenburg, Sweden. His main research interest focuses on traffic network modeling and optimization.
\endbio

\bio{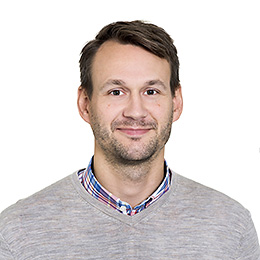}
\noindent \textbf{Anders Lindman} received his Ph.D. (2017) in Computational Materials Physics from Chalmers University of Technology (Sweden). He joined Volvo Car Corporation in 2018, where he has been active in the areas of Data \& Analytics, Vehicle Energy Efficiency and Future Mobility.
\endbio

\end{document}